\documentclass[english,aps,manuscript]{revtex4}
\usepackage[T1]{fontenc}
\usepackage[latin9]{inputenc}
\usepackage{amsmath,stackrel}
\usepackage{amssymb}
\usepackage{stackrel}
\usepackage{graphicx}
 \usepackage{ulem}
 \usepackage{color}
 \usepackage[dvipsnames]{xcolor}
 \usepackage{cancel}

\makeatletter
\@ifundefined{textcolor}{}
{%
 \definecolor{BLACK}{gray}{0}
 \definecolor{WHITE}{gray}{1}
 \definecolor{RED}{rgb}{1,0,0}
 \definecolor{GREEN}{rgb}{0,1,0}
 \definecolor{BLUE}{rgb}{0,0,1}
 \definecolor{CYAN}{cmyk}{1,0,0,0}
 \definecolor{MAGENTA}{cmyk}{0,1,0,0}
 \definecolor{YELLOW}{cmyk}{0,0,1,0}

\definecolor{nred}{RGB}{224,0,0}
\definecolor{nblue}  {RGB}{28,130,185}
\definecolor{dgreen} {RGB}{38,238,21}
\definecolor{norange}{RGB}{230,120,20}

}

\makeatother

\usepackage{babel}
\begin{document}
\title{Mass generation at a fixed point: A Functional Renormalization Group Study of the tricritical O($N$) model in $d=3$ and $N=\infty$ }

\author{Shunsuke Yabunaka}

\affiliation{Advanced Science Research Center, Japan Atomic Energy Agency, Tokai, 319-1195, Japan}

\author{Bertrand Delamotte}

\affiliation{Sorbonne Universit\'e, CNRS, Laboratoire de Physique Th\'eorique de la Mati\`ere Condens\'ee, LPTMC, F-75005 Paris, France}
\date{\today}
\begin{abstract}
  Renormalization group (RG) fixed points are commonly associated with scale invariance and a divergent correlation length. We show that this connection can fail in the tricritical $O(N)$ model in three dimensions in the limit $N\to\infty$. Revisiting the line of fixed points identified by Bardeen, Moshe, and Bander, we use the functional renormalization group to clarify the mechanism leading to mass generation at its singular endpoint (the BMB fixed point). We demonstrate that the generated mass is nonuniversal and originates from the nonanalytic structure of the effective potential.
We  show that  the critical exponent $\nu$ which takes the value $\nu = 1/2$ along the regular part of the BMB line,  that is, for $0 \leq \lambda < \lambda_{\rm BMB}$,  jumps to $\nu = 1/3$ on the singular part of this line with the BMB FP, corresponding to $\lambda = \lambda_{\rm BMB}$, being the pivotal point between these two regimes. We also show how its singular potential emerges dynamically along the renormalization flow.
\end{abstract}
\maketitle

\section{Introduction}

Second-order phase transitions are commonly associated with renormalization group (RG) fixed points (FPs) and the emergence of scale invariance at distances large compared with the ultraviolet cutoff of the model (e.g., the lattice spacing in lattice systems with short-range interactions). Scale invariance is typically rooted in the divergence of the correlation length, or equivalently, in high-energy terminology, in the vanishing of the renormalized mass. Since the mass is given by the curvature of the effective potential at vanishing field, criticality is expected to imply a small-field behavior of the form
\begin{equation}
U_{\rm eff}(\phi) \underset{\phi\to0}{\propto} \phi^{\delta+1}
\quad {\rm with} \quad\delta > 1
\end{equation}
such that its curvature is vanishing at $\phi=0$. This scenario holds for most standard models: the existence of an RG fixed point entails dilatation invariance at large scales and a divergent correlation length.
In this work, we reexamine this connection, which is not as general as commonly assumed. In particular, a FP does not necessarily correspond to a massless theory. We revisit this issue in the tricritical 
O($N$) model in three dimensions in the limit $N\to\infty$, where this paradox was identified long ago by Bardeen, Moshe, and Bander (BMB) \citep{bardeen1984spontaneous}. Our main goal is to demonstrate that the functional renormalization group (FRG) provides a simple and comprehensive framework to understand this phenomenon. Similar phenomena have been found also in supersymmetric $O(N)$ models \cite{PhysRevD.86.105006}.

Bardeen, Moshe, and Bander  showed that the $O(N)$ model with interaction 
$\lambda (\boldsymbol{\varphi}^2)^3$, which describes the tricritical behavior of the model, displays several remarkable features in $d=3$ in the large-$N$ limit. They uncovered a continuous line of FPs parametrized by $\lambda$, hereafter referred to as the BMB line. This line originates at the Gaussian FP at $\lambda=0$ and terminates at a special FP, which we denote as the BMB fixed point, corresponding to a particular positive value $\lambda_{\rm BMB}$. All these FPs are twice infrared unstable and are therefore tricritical. Although interacting, they share identical critical exponents, with the notable exception of the endpoint BMB fixed point \citep{david1985study,David2}. These FPs provide a rare example of asymptotically safe scalar theories \citep{david1985study,David2,litim2018asymptotic,litim2017fixed}. 

The BMB fixed point itself is singular in the sense that its effective potential develops a nonanalytic behavior at small field \cite{david1985study,David2,litim2018asymptotic,litim2017fixed}. As said above, dilatation invariance is broken at this FP and we show how the breaking of scale invariance at criticality is related to the existence of a singularity at vanishing field of the BMB fixed point potential. Notice that BMB interpreted the emergence of a massless bound state as a consequence of the spontaneous breaking of dilatation symmetry.

In Refs.~\citep{yabunaka2017surprises,yabunaka2018might}, we showed that the BMB line admits a finite-$N$ extension, albeit in a subtle manner: it exists only when the limit $N \to \infty$ is taken along the hyperbolae $d = 3 - \alpha/N$ in the $(d,N)$ plane, rather than at fixed dimension $d=3$. The parameter $\alpha$ is real and varies on a finite interval, see the following. Our analysis further revealed that the conventional BMB line represents only one half of the full line of FPs. The second half, which extends the line from the BMB fixed point, consists of singular FPs whose effective potentials have a cusp at a finite field value. At finite and large $N$, these cusps are smoothed into boundary layers.

In this work, we demonstrate that the functional renormalization group provides a natural and powerful framework to address these issues, and in particular to clarify the mechanism leading to mass generation at the BMB fixed point. Within this approach, we show that the generated mass is not universal: by appropriately tuning the bare action, one can obtain an arbitrary mass scale at the BMB fixed point which is related to the singularity of its FP potential at $\phi=0$. 

We also show that  the critical exponent $\nu$ which takes the value $\nu = 1/2$ along the regular part of the BMB line,  that is, for $0 \leq \lambda < \lambda_{\rm BMB}$,  jumps to $\nu = 1/3$ on the singular part of this line with the BMB FP, corresponding to $\lambda = \lambda_{\rm BMB}$, being the pivotal point between these two regimes.  Finally, we explicitly demonstrate how the nonanalytic structure of the BMB effective potential emerges dynamically along the RG flow, thereby elucidating the origin of its singular behavior.

\section{nonperturbative renormalization group (NPRG)}
A central tool in our analysis is the Wilsonian functional and nonperturbative renormalization group (NPRG). As we show below, it not only provides an efficient framework to compute FPs that are inaccessible to conventional perturbative or $1/N$ approaches, but also enables the study of phenomena that cannot even be properly formulated within a standard perturbative expansion \citep{yabunaka2017surprises,yabunaka2018might,tissier08,tissier10,TissierPhysRevB.74.214419,CanetPhysRevE.93.063101,CanetPhysRevE.84.061128,GredatPhysRevE.89.010102}. 
This includes, in particular, FPs whose effective potentials develop nonanalytic structures such as cusps or boundary layers. We therefore briefly introduce the NPRG formalism.

The NPRG is based on Wilson's idea of integrating fluctuations progressively, from short to long distances 
\citep{PhysRevB.4.3174}. 
In its modern formulation, it is implemented at the level of the scale-dependent Gibbs free energy $\Gamma_k$, which is the generating functional of one-particle irreducible correlation functions \citep{wetterich91,wetterich93b,Ellwanger,Morris94,delamotte2012}.
 One defines a one-parameter family of models indexed by a momentum scale $k$, with partition functions ${\cal Z}_k$, such that only fluctuations with wavenumbers $|q|>k$ are integrated out in ${\cal Z}_k$. The slow modes ($|q|<k$) are suppressed by adding to the original Hamiltonian $H$ a quadratic (mass-like) regulator term:
\begin{equation}
{\cal Z}_{k}[\boldsymbol{J}]=\int D\boldsymbol{\varphi}\,
\exp\!\left(-H[\boldsymbol{\varphi}]-\Delta H_{k}[\boldsymbol{\varphi}]
+\boldsymbol{J}\cdot\boldsymbol{\varphi}\right),
\end{equation}
where $\boldsymbol{\varphi}$ is an $N$-component scalar field, $H$ is a general $O(N)$-invariant Hamiltonian and 
$\boldsymbol{J}\cdot\boldsymbol{\varphi}
=\int_{x}J_{i}(x)\varphi_{i}(x)$. The regulator term is chosen as
\begin{equation}
\Delta H_{k}[\boldsymbol{\varphi}]
=\frac{1}{2}\int_{q}R_{k}(q^{2})\,
\varphi_{i}(q)\varphi_{i}(-q),
\end{equation}
with, for instance,
\begin{equation}
R_{k}(q^{2})=\bar{Z}_{k}(k^{2}-q^{2})\theta(k^{2}-q^{2}),
\label{eq:Litim-reg}
\end{equation}
where $\theta$ is the Heaviside function and $\bar{Z}_{k}$ denotes the running field renormalization. 

The scale-dependent Gibbs free energy $\Gamma_{k}[\boldsymbol{\phi}]$, with $\phi_{i}=\langle\varphi_{i}\rangle$, is defined as a modified Legendre transform of $\log {\cal Z}_{k}[\boldsymbol{J}]$:
\begin{equation}
\Gamma_{k}[\boldsymbol{\phi}]
+\log{\cal Z}_{k}[\boldsymbol{J}]
=\boldsymbol{J}\cdot\boldsymbol{\phi}
-\frac{1}{2}\int_{q}R_{k}(q^{2})\phi_{i}(q)\phi_{i}(-q).
\label{legendre}
\end{equation}
Its exact flow equation, derived by Wetterich \citep{wetterich93b}, reads
\begin{equation}
\partial_{t}\Gamma_{k}[\boldsymbol{\phi}]
=\frac{1}{2}\mathrm{Tr}
\left[
\partial_{t}R_{k}\,
\left(\Gamma_{k}^{(2)}+R_{k}\right)^{-1}
\right],
\label{flow}
\end{equation}
where $t=\log(k/\Lambda)$, $\Lambda$ is the ultraviolet cutoff of the model (e.g., the inverse lattice spacing in lattice systems), and $\Gamma_{k}^{(2)}$ denotes the matrix of second functional derivatives of $\Gamma_k$ with respect to the fields. The trace involves both momentum (or coordinate) integration and summation over internal indices.

From Eq.~(\ref{legendre}) and the properties of the regulator, one can show that $\Gamma_{k=\Lambda}=H$, since no fluctuations are included at the ultraviolet scale, while $\Gamma_{k=0}=\Gamma$, where all fluctuations have been integrated out. Thus, for any finite scale $k<\Lambda$, $\Gamma_k$ continuously interpolates between the microscopic Hamiltonian $H$ and the full Gibbs free energy $\Gamma$.

For the models considered here, the exact flow equation~(\ref{flow}) cannot be solved analytically and one must resort to approximations . A powerful nonperturbative approximation scheme consists in performing a derivative expansion of $\Gamma_k[\boldsymbol{\phi}]$ in powers of $\nabla \boldsymbol{\phi}$ \citep{berges2002,delamotte04,canet03,canet04,PhysRevLett.123.240604,PhysRevE.106.024111}. 
At lowest order, known as the local potential approximation (LPA), the effective action is approximated by
\begin{equation}
\Gamma_{k}^{\text{LPA}}[\boldsymbol{\phi}]
=
\int_{x}
\left(
\frac{1}{2}(\nabla \phi_i)^2
+
U_k(\rho)
\right),
\label{LPA}
\end{equation}
where $\rho=\phi_i \phi_i/2$. In the LPA ansatz~(\ref{LPA}), the prefactor of the gradient term, $1/2(\nabla \phi_i)^2$, is fixed to unity rather than promoted to a running wavefunction renormalization factor $\bar{Z}_k$.  As a consequence, field renormalization effects are neglected at this level of approximation, and the anomalous dimension $\eta$ vanishes identically within the LPA.

At criticality, the system becomes scale invariant and the RG flow approaches a FP, which can only be identified in terms of dimensionless quantities. We therefore introduce renormalized, dimensionless variables by rescaling coordinates and fields according to
\begin{equation}
\tilde{x}=k x,
\qquad
\tilde{\boldsymbol{\phi}}(\tilde{x})
=
v_d^{-1/2}
k^{(2-d)/2}
\boldsymbol{\phi}(x),
\label{dimensionless}
\end{equation}
where
\begin{equation}
v_d^{-1}
=
2^{d-1} d \pi^{d/2} \Gamma\!\left(\frac{d}{2}\right).
\end{equation}
has been introduced for convenience in the definition of $\tilde{\boldsymbol{\phi}}$, Eq.~\eqref{dimensionless}.
Since at LPA $\bar Z_k=1$, the dimensionless potential is defined from its canonical dimension as
\begin{equation}
\tilde{U}_k(\tilde{\rho})
=
v_d^{-1}
k^{-d}
U_k(\rho)
\end{equation}
where the $v_d$ factor has again been introduced for convenience.
In practice, the flow of the potential is obtained as follows. First, for a uniform field configuration, the potential is defined by
\begin{equation}
U_k(\boldsymbol{\phi})
=
\frac{1}{\Omega}
\Gamma_k[\boldsymbol{\phi}],
\end{equation}
where $\Omega$ denotes the volume of the system. Acting with $\partial_t$ on this definition and using Eq.~(\ref{flow}), together with the LPA ansatz~(\ref{LPA}) to evaluate $\Gamma_k^{(2)}$, one obtains a closed flow equation for $U_k$. Expressed in dimensionless form and using the regulator~(\ref{eq:Litim-reg}), it reads
\begin{equation}
\partial_t \tilde{U}_t(\tilde{\phi})
=
-d\,\tilde{U}_t(\tilde{\phi})
+
\frac{1}{2}(d-2)\tilde{\phi}\,\tilde{U}_t'(\tilde{\phi})
+
(N-1)
\frac{\tilde{\phi}}
{\tilde{\phi}+\tilde{U}_t'(\tilde{\phi})}
+
\frac{1}
{1+\tilde{U}_t''(\tilde{\phi})},
\label{eq:flow-LPA}
\end{equation}
where $\tilde{\phi}=|\tilde{\boldsymbol{\phi}}|$.

It is convenient to introduce variables that remain finite in the large-$N$ limit:
\begin{equation}
\bar{\phi}
=
\frac{\tilde{\phi}}{\sqrt{N}},
\qquad
\bar{U}_t(\bar{\phi})
=
\frac{\tilde{U}_t(\tilde{\phi})}{N}.
\label{eq:bar-tilde}
\end{equation}
In terms of these rescaled quantities, the FP equation associated with~(\ref{eq:flow-LPA}) becomes
\begin{equation}
0
=
-d\,\bar{U}
+
\frac{1}{2}(d-2)\bar{\phi}\,\bar{U}'
+
\left(1-\frac{1}{N}\right)
\frac{\bar{\phi}}
{\bar{\phi}+\bar{U}'}
+
\frac{1}{N}
\frac{1}
{1+\bar{U}''}.
\label{eq:flow-LPA scaled in N}
\end{equation}
In the limit $N\to\infty$, this equation simplifies to
\begin{equation}
0
=
-d\,\bar{U}
+
\frac{1}{2}(d-2)\bar{\phi}\,\bar{U}'
+
\frac{\bar{\phi}}
{\bar{\phi}+\bar{U}'}
\label{eq:LargeNflow}
\end{equation}
provided the last term in Eq.~\eqref{eq:flow-LPA scaled in N} is negligible, see below.

\section{The BMB line in $d=3$ and at $N=\infty$ }

\begin{figure}
\includegraphics[scale=0.2]{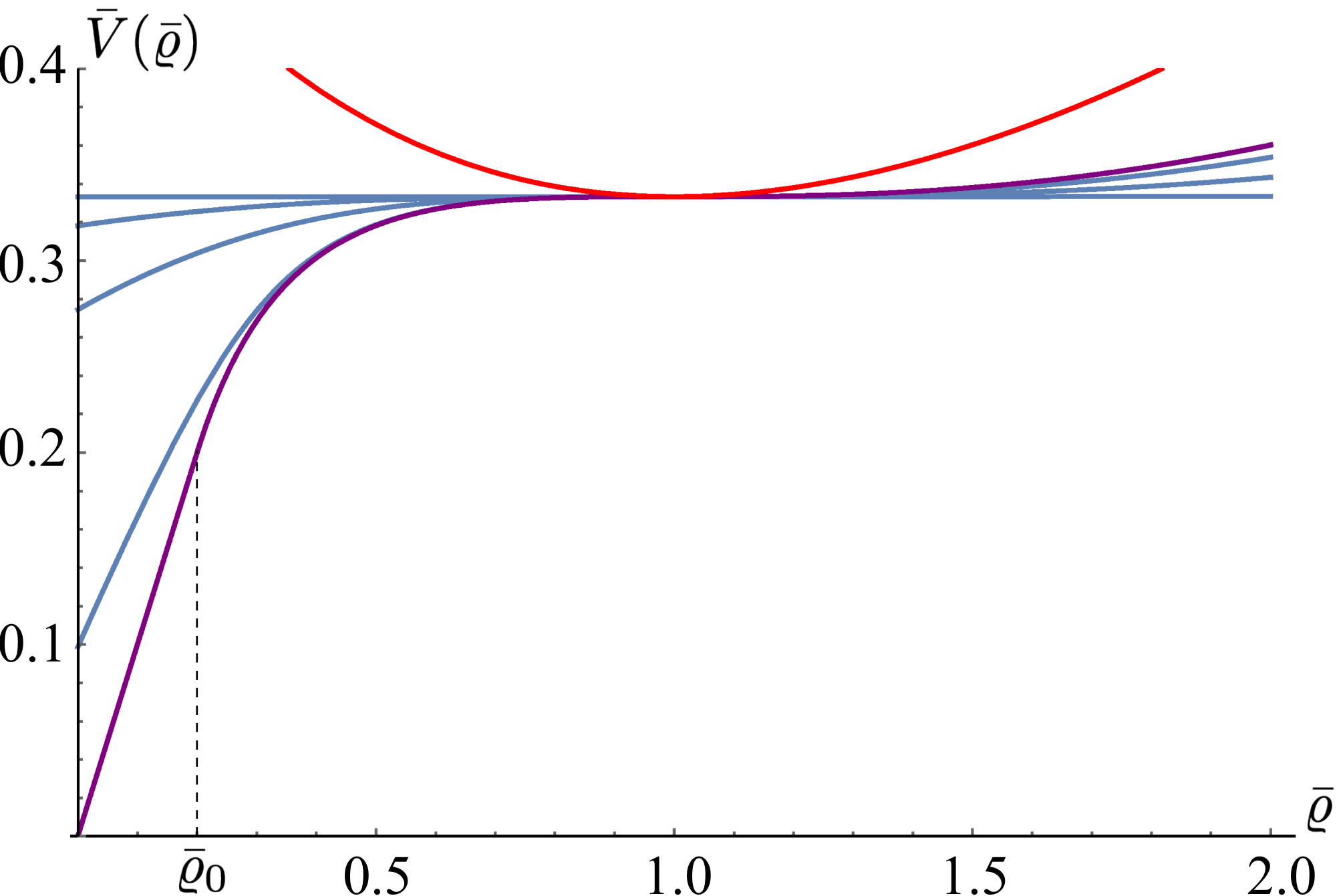} \includegraphics[scale=0.27]{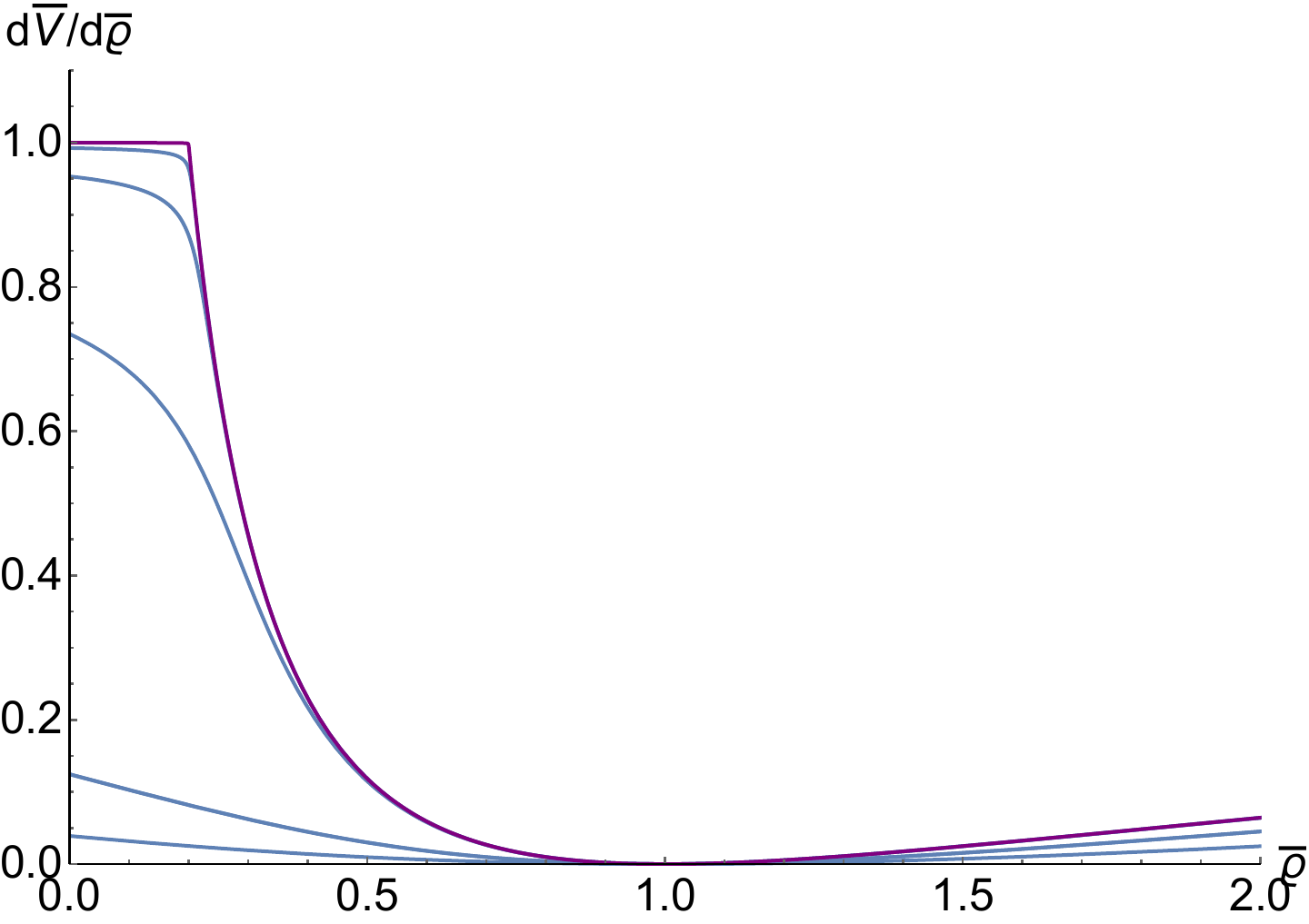}
\caption{Wilson-Polchinski parameterization: Plot of the potentials $\bar{V}(\bar{\varrho})$  and of $\bar{V}'(\bar{\varrho})$ for some 
of the tricritical FPs ${\cal A}(\tau)$ of the BMB line (blue) together
with the Wilson-Fisher FP (red). The BMB FP is the endpoint of the
BMB line (purple). All these potentials are given by Eq. (\ref{solrho}).
The Gaussian FP $G$ corresponds to the horizontal line. The BMB FP
potential shows a discontinuity in its second derivative at $\bar{\varrho}_{0}$.} 
\label{fig:FP potentials on the BMB line WP}
\end{figure}

\begin{figure}
\includegraphics[scale=0.5]{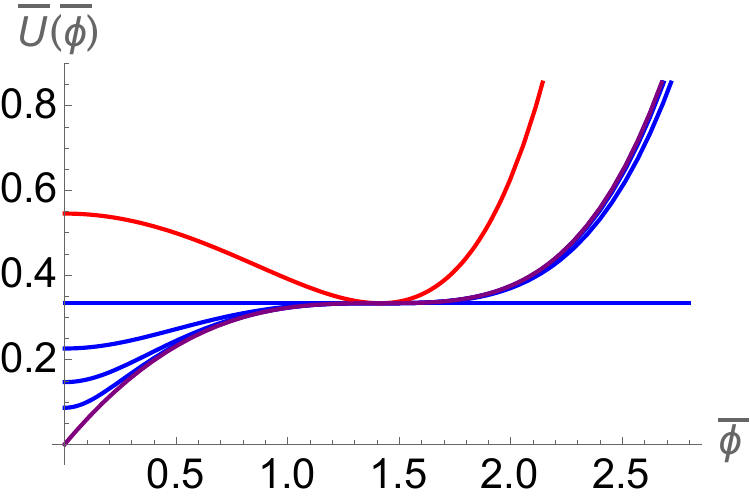}\ \ \ \ \ \ \ \ \ \includegraphics[scale=0.5]{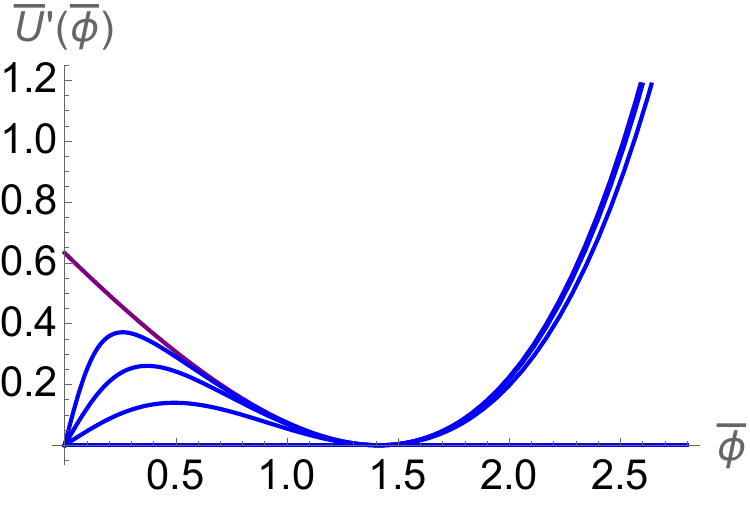}

\includegraphics[scale=0.45]{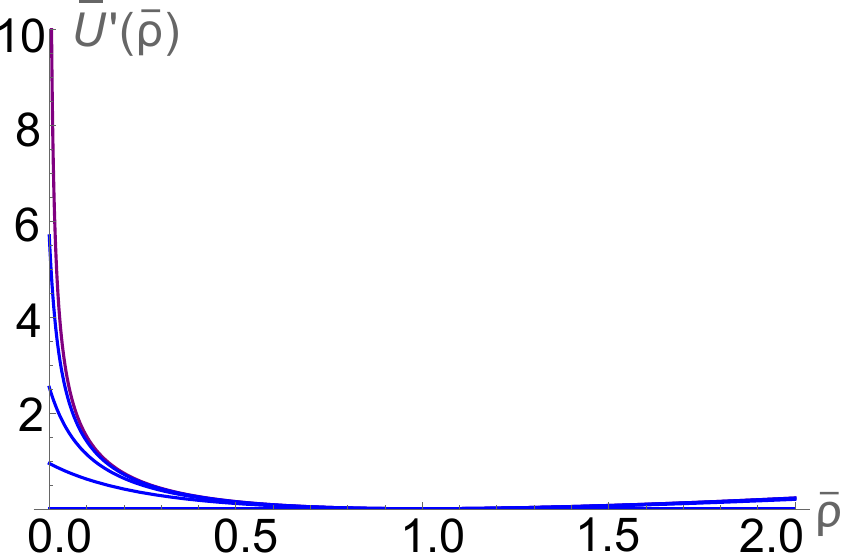}
\caption{  Wetterich parameterization. 
Upper left: FP potentials $\bar U(\bar\phi)$ corresponding to the tricritical line ${\cal A}(\tau)$ with $0\le\tau<\tau_{\rm BMB}$ (blue curves), together with the Wilson-Fisher FP (red) and the BMB FP (purple), which is the endpoint of the BMB line. The Gaussian FP is shown as the horizontal line. The tricritical potentials are obtained from Eq.~(\ref{solrho}) (derived in the Wilson-Polchinski formulation of the LPA flow).
Upper right: Derivatives $\bar U'(\bar\phi)$ of the FP potentials shown in the upper-left panel. For all FPs along the BMB line, except the BMB FP, the potentials behave as $\bar U(\bar\phi)\sim \bar\phi^{2}$ near the origin, implying $ \partial_{\bar\phi}\bar{U}(\bar\phi=0)=0$. 
Lower: Derivatives $\partial_{\bar\rho}\bar U(\bar\rho)$ of the same FP potentials plotted as functions of $\bar\rho=\bar\phi^{2}/2$. While $\partial_{\bar\rho}\bar U(\bar\rho)$ remains finite at $\bar\rho=0$ for all regular FPs, it diverges at the origin for the BMB FP, $\bar U_{\rm BMB}'(\bar\rho=0)=\infty$, which is therefore  nonanalytic at $\bar\rho=0$.}
\label{fig:FPs along the BMB line}
\end{figure}

For the analysis of the large-$N$ limit, it is particularly convenient to perform a change of variables that considerably simplifies the structure of the flow equation. This transformation is equivalent to working with the Wilson-Polchinski formulation of the RG instead of the Wetterich flow for $\Gamma_k$. Following Ref.~\citep{Morris}, we introduce the potential  $\tilde V(\tilde\varrho)$ defined by
\begin{equation}
\tilde{V}_{t}(\tilde{\varrho})
=
\tilde{U}_{t}(\tilde{\rho})
+
\frac{1}{2}
\left(
\tilde{\phi}_i - \tilde{\Phi}_i
\right)^2,
\label{eq:transformation-WP-W}
\end{equation}
with $\tilde{\varrho}=\tilde{\Phi}_i \tilde{\Phi}_i/2=\tilde{\Phi}^2/2$ and
\begin{equation}
\tilde{\phi}_i-\tilde{\Phi}_i
=
-\tilde{\Phi}_i \tilde{V}'_{t}(\tilde{\varrho})
=
-\tilde{\phi}_i \tilde{U}'_{t}(\tilde{\rho}).\label{eq:transform-WP-W-2}
\end{equation}

As before, we introduce rescaled variables
\begin{equation}
\bar{\varrho}=\frac{\tilde{\varrho}}{N},
\qquad
\bar V_t(\bar{\varrho})=\frac{\tilde V_t(\tilde{\varrho})}{N}.
\end{equation}
 The corresponding fixed-point (FP) equation reads \citep{litim2018asymptotic,Morris,yabunaka2018might}
\begin{equation}
0
=
1
-
d\,\bar V
+
(d-2)\bar{\varrho}\bar V'
+
2\bar{\varrho}(\bar V')^2
-
\bar V'
-
\frac{2}{N}\bar{\varrho}\bar V'',
\label{flow-LPA-WP}
\end{equation}
where primes denote derivatives with respect to $\bar{\varrho}$.

In the standard large-$N$ limit, one assumes that $\bar V(\bar{\varrho})$ remains regular for all $\bar{\varrho}$. Under this assumption, the last term in Eq.~(\ref{flow-LPA-WP}), can be discarded when $N\to\infty$. In $d=3$ and at $N=\infty$, the resulting equation admits infinitely many solutions \citep{bardeen1984spontaneous,Omid,litim2018asymptotic,litim2017fixed}. For the physical solutions of interest here, they are given implicitly by
\begin{equation}
\bar{\varrho}_{\pm}
=
1
+
\frac{\bar V'\left(\frac{5}{2}-\bar V'\right)}
{(1-\bar V')^2}
+
\frac{\frac{3}{2}\arcsin\sqrt{\bar V'} \pm \sqrt{2/\tau}}
{(\bar V')^{-1/2}(1-\bar V')^{5/2}},
\label{solrho}
\end{equation}
where $\bar{\varrho}_+(\bar V')$ and $\bar{\varrho}_-(\bar V')$ correspond to the branches $\bar{\varrho}>1$ and $\bar{\varrho}<1$, respectively, and $\tau$ is an integration constant.

A detailed analysis of Eq.~(\ref{solrho}) shows that:

(i) the Gaussian FP, defined by $\bar V'(\bar{\varrho})=0$, is obtained for $\tau=0$;

(ii) a well-defined solution $\bar V(\bar{\varrho})$ exists for all
\begin{equation}
\tau \in [0,\tau_{\rm BMB}], 
\qquad
\tau_{\rm BMB}=\frac{32}{(3\pi)^2},
\end{equation}
which defines the BMB line of fixed points, denoted ${\cal A}(\tau)$. The endpoint, $\tau=\tau_{\rm BMB}$, corresponds to the BMB FP \citep{bardeen1984spontaneous,david1985study,David2,litim2018asymptotic,litim2017fixed};

(iii) for $\tau>\tau_{\rm BMB}$, the solutions are no longer defined over the full interval $\bar{\varrho}\in[0,\infty[$ \citep{litim2018asymptotic} and are therefore non physical;

(iv) an isolated solution exists in the limit $\sqrt{2/\tau}\to0$, corresponding to the Wilson-Fisher FP of the $O(N=\infty)$ model (an analytic continuation is required for $\bar V'<0$).

All FPs with $\tau\in[0,\tau_{\rm BMB}[$ are twice infrared unstable and are therefore tricritical and their potentials are regular for all $\bar{\varrho}$. As $\tau\to\tau_{\rm BMB}$, the FP potentials approach a limiting shape whose second derivative becomes singular at a finite value $\bar{\varrho}_0$ (see Fig.~\ref{fig:FP potentials on the BMB line WP}).

Using Eqs.~(\ref{eq:bar-tilde}) and (\ref{eq:transformation-WP-W}), one recovers the corresponding FP potentials in the Wetterich formulation \citep{wetterich93b,berges2002}. The tricritical line can be written implicitly as
\begin{equation}
\bar{\rho}
=
1
+
\sqrt{\bar U'(\bar\rho)}
\left(
F(\bar U'(\bar\rho))+c
\right),
\label{eq:tricr-line}
\end{equation}
with
\begin{equation}
F(\bar U'(\bar\rho))
=
\frac{1}{2}
\frac{\sqrt{\bar U'(\bar\rho)}}{1+\bar U'(\bar\rho)}
+
\frac{3}{2}
\arctan\sqrt{\bar U'(\bar\rho)},
\label{eq:Fu'}
\end{equation}
where $\bar U'(\bar\rho)$ is the derivative of the tricritical FP potential indexed by $\tau$ and $c=2/\sqrt{\tau}$. The limits $c=\infty$ and $c_{BMB}=3\pi/4$ correspond to the Gaussian and BMB fixed points, respectively.

In the Wetterich parameterization, the derivative BMB fixed point potential develops a cusp, because $\bar U(\bar\rho)\simeq {\rm const} \times|\bar\phi|$ around  vanishing field $\bar\phi=0$ \citep{litim2017fixed}. This singularity has been interpreted as signaling the spontaneous breaking of scale invariance, and it has been conjectured that the physical mass at the BMB FP is a free parameter \citep{litim2017fixed,litim2018asymptotic}. However, the relation between this mass and the choice of bare potential at the ultraviolet scale remains unclear. We address this question in Sec.~\ref{sec:Flow-of-dimensionfull}.

\section{The singular BMB line at $N=\infty$ and $d=3$}

\begin{figure}
\begin{centering}
\includegraphics[width=0.4\columnwidth]{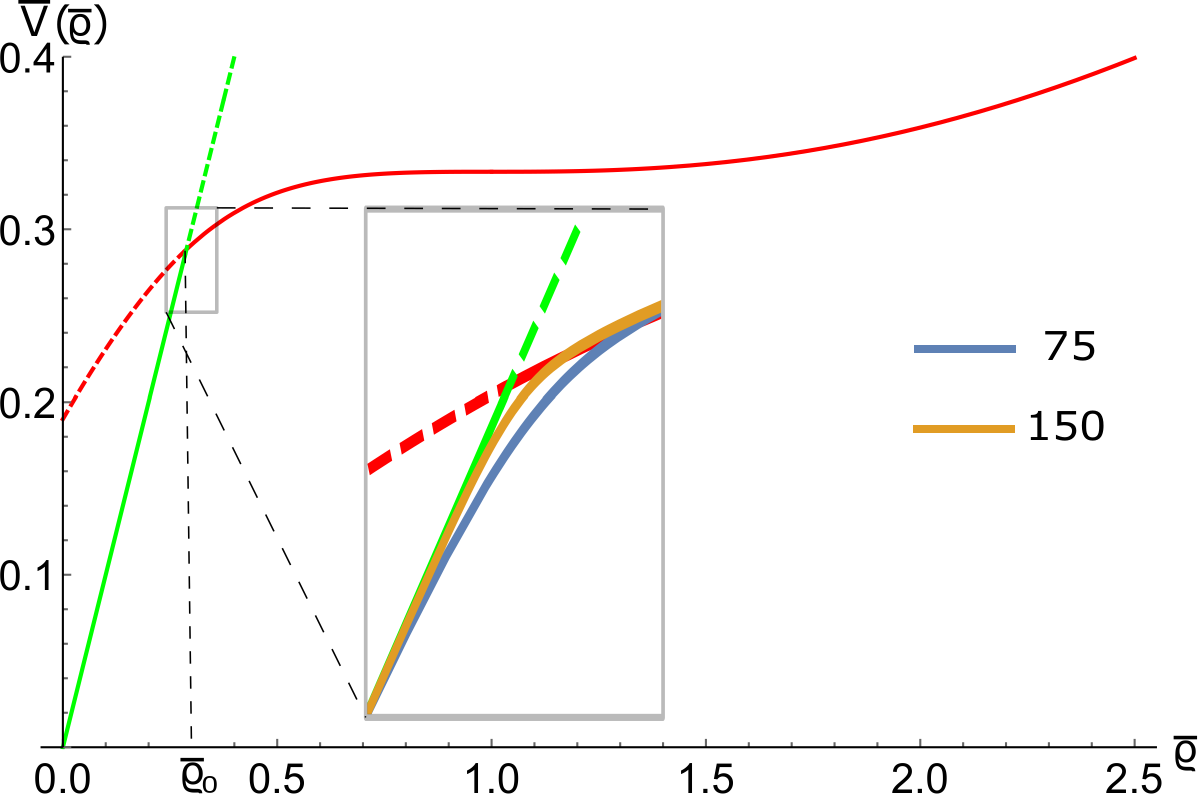}
\par\end{centering}
\caption{$N=\infty$ and $d=3$: Construction of the singular Wilson-Polchinski fixed-point potential of the $S{\cal A}(\tau=0.33)$ FP. 
The regular fixed-point potential ${\cal A}(\tau=0.33)$ obtained from 
Eq.~(\ref{solrho}) is shown by the red curves (solid and dashed). 
The green curves (solid and dashed) correspond to the linear solution 
$\bar V(\bar{\varrho})=\bar{\varrho}$, which represents the high-temperature (Gaussian) fixed point. 
The singular fixed-point potential of $S{\cal A}(\tau=0.33)$ is constructed 
by replacing the small-$\bar{\varrho}$ part of the potential of ${\cal A}(\tau=0.33)$ 
with the linear branch and matching the two solutions continuously 
at $\bar{\varrho}_{0}(\tau=0.33)$. 
The resulting potential is given by the solid green and red curves, 
which meet at $\bar{\varrho}_{0}(\tau=0.33)$ and form a cusp. 
Inset: Enlargement of the cusp region, showing its smoothing at finite $N$ due to the emergence of a boundary layer.
}
\label{sing-pot-polchinski}
\end{figure}

As emphasized above, the regular BMB line ${\cal A}(\tau)$ of fixed points must be supplemented by a singular counterpart \citep{Fleming2020,yabunaka2018might}. We briefly recall how these additional fixed points arise. In this section, we work within the Wilson-Polchinski formulation of the RG flow, which proves technically more convenient.

Because the potentials of interest are singular, the functional space in which Eq.~(\ref{flow-LPA-WP}) is solved at $N=\infty$ must be enlarged. We restrict ourselves to solutions displaying a cusp at an isolated value of $\bar{\varrho}$. Such solutions can be constructed by matching piecewise regular solutions of Eq.~(\ref{flow-LPA-WP}) in which the $1/N$ term is discarded, as appropriate in the strict large-$N$ limit.

As discussed above, the  derivative of the BMB fixed-point potential is discontinuous at a finite value $\bar{\varrho}_0$, see Fig. \ref{fig:FP potentials on the BMB line WP}.  The BMB potential is in fact composed of two distinct parts: a linear branch,
\begin{equation}
   \bar V(\bar{\varrho}) = \bar{\varrho}, 
\end{equation}
valid for $\bar{\varrho}\le \bar{\varrho}_0$, and a nonlinear branch with $\bar V''(\bar{\varrho})\neq 0$ for $\bar{\varrho}>\bar{\varrho}_0$. The matching point $\bar{\varrho}_0$ is determined by the continuity of the first derivative $\bar V'(\bar{\varrho})$ see Fig. \ref{fig:FP potentials on the BMB line WP}.

This structure naturally suggests the construction of a ``singular copy'' of the BMB line. Consider any regular fixed point ${\cal A}(\tau)$ along the BMB line. One may replace the small-$\bar{\varrho}$ portion of its potential by the linear solution $\bar V(\bar{\varrho})=\bar{\varrho}$ -- which itself satisfies the fixed-point equation~(\ref{flow-LPA-WP}) at $N=\infty$ -- up to a matching point $\bar{\varrho}_0(\tau)$ where the linear and nonlinear branches join continuously, see Fig.~\ref{sing-pot-polchinski}. 

The resulting potential solves Eq.~(\ref{flow-LPA-WP}) for all $\bar{\varrho}\neq \bar{\varrho}_0(\tau)$ and therefore defines a fixed point. Its distinctive feature is the presence of a cusp at $\bar{\varrho}_0(\tau)$. We denote this family of singular fixed points by $S{\cal A}(\tau)$, where $S$ stands for ``singular''. 

We thus conclude that the conventional BMB line ${\cal A}(\tau)$ represents only one half of the full line of fixed points at $N=\infty$. In this construction, the BMB fixed point plays a pivotal role, as the entire singular branch can be viewed as a continuous deformation of its potential.

\section{Finite-$N$ extension of the BMB line and Singular BMB line}
\label{finite-N-extension}
\begin{figure}
\includegraphics{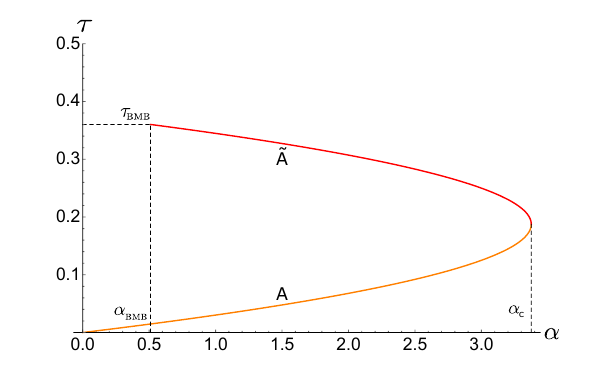}\caption{Plot of the two solutions $\tau_{1}(\alpha)$ (orange curve) and $\tau_{2}(\alpha)$ (red curve),  corresponding to the FPs $A$ and $\tilde{A}$, as a function of $\alpha$.}
\label{fig: alpha}
\end{figure}

We showed in \citep{Fleming2020,Yabunaka2022} that the BMB line has an intriguing finite $N$ origin. We summarize below what it consists in because this is needed for the following.
As shown in \citep{Fleming2020,Yabunaka2022}, the limit $N\to\infty$ and $d\to3$ must be taken at fixed $\alpha=\epsilon N$ with $\epsilon=3-d$, that is, along hyperbolae of the $(d,N)$-plane: $N=\alpha/(3-d)$. Proceeding this way, it is possible to show that to each FP ${\cal A}(\tau)$ with $\tau\in[0,\tau_{\text{ BMB}}]$ on the BMB line, there is one FP at finite $N$ 
that converges to ${\cal A}(\tau)$ when
$N\to\infty$. The relation between admissible values
of $\tau$, that is, values for which a FP on the BMB line exists,
and admissible values of $\alpha$ where 
a FP exists at finite $N$ is given at LPA by:
\begin{equation}
\alpha-36\tau+96\tau^{2}=0.
\label{eq:alpha1}
\end{equation}
This equation has two solutions $\tau_{1}(\alpha)$ and $\tau_{2}(\alpha)$
that we choose such that $\tau_{1}(\alpha)\le\tau_{2}(\alpha)$ as
shown in Fig. \ref{fig: alpha}. This implies that to each value of $\alpha$ correspond two FPs on the BMB line that we call $A=A(\tau_{1}(\alpha))$ and $\tilde{A}=\tilde{A}(\tau_{2}(\alpha))$ and that are
the limits of two different FPs existing at finite $N$ for each admissible value of $\alpha$. Thus, the set of ${\cal A}(\tau)$ FPs at $N=\infty$ is the union of the sets of $A(\tau_{1}(\alpha))$ and $\tilde A(\tau_{2}(\alpha))$ FPs. In our previous papers \citep{yabunaka2017surprises,yabunaka2018might,Yabunaka2022}, at finite $N$, $A$ and $\tilde{A}$ are respectively called $A_2$ and $\tilde{A}_3$, where the subscript refers to the number of relevant directions around the FP.

The same scenario takes place for the singular FPs $S{\cal A}(\tau)$ belonging to the singular part of the BMB line: they are the limit of two other FPs existing at finite $N$ when they are followed along the same  hyperbolae as before. The relationship between $\tau$ and $\alpha$ is again given by Eq.~ \eqref{eq:alpha1} and the finite $N$ FPs are smooth but show a boundary layer that becomes a singularity when $N=\infty$.

In other words, at finite $N$, there should
exist a boundary layer around the point $\bar{\varrho}_{0}(\tau)$, see Fig.~\ref{sing-pot-polchinski},
such that inside the layer the potential varies smoothly -- but abruptly
-- in order to connect the linear part of the potential for $\bar{\varrho}<\bar{\varrho}_{0}(\tau)$
to the nontrivial part of the potential for $\bar{\varrho}>\bar{\varrho}_{0}(\tau)$.
Moreover, the boundary layer should be sufficiently thin such that
$\bar{V}''$ varies as $N$ at large $N$ in such a way that inside
the layer it compensates the $1/N$ factor in front of it in Eq. (\ref{flow-LPA-WP}).
This is achieved by a layer of typical width $1/N$. In this case,
all the terms of Eq. (\ref{flow-LPA-WP}), including the last
one, must be retained in the large $N$ limit because they are all
of the same order in $N$. 

\section{Eigenvalues of the linearized RG flow around regular and singular FPs on the BMB line }
\label{sec:linearized} 

\begin{figure}
\includegraphics[scale=0.5]{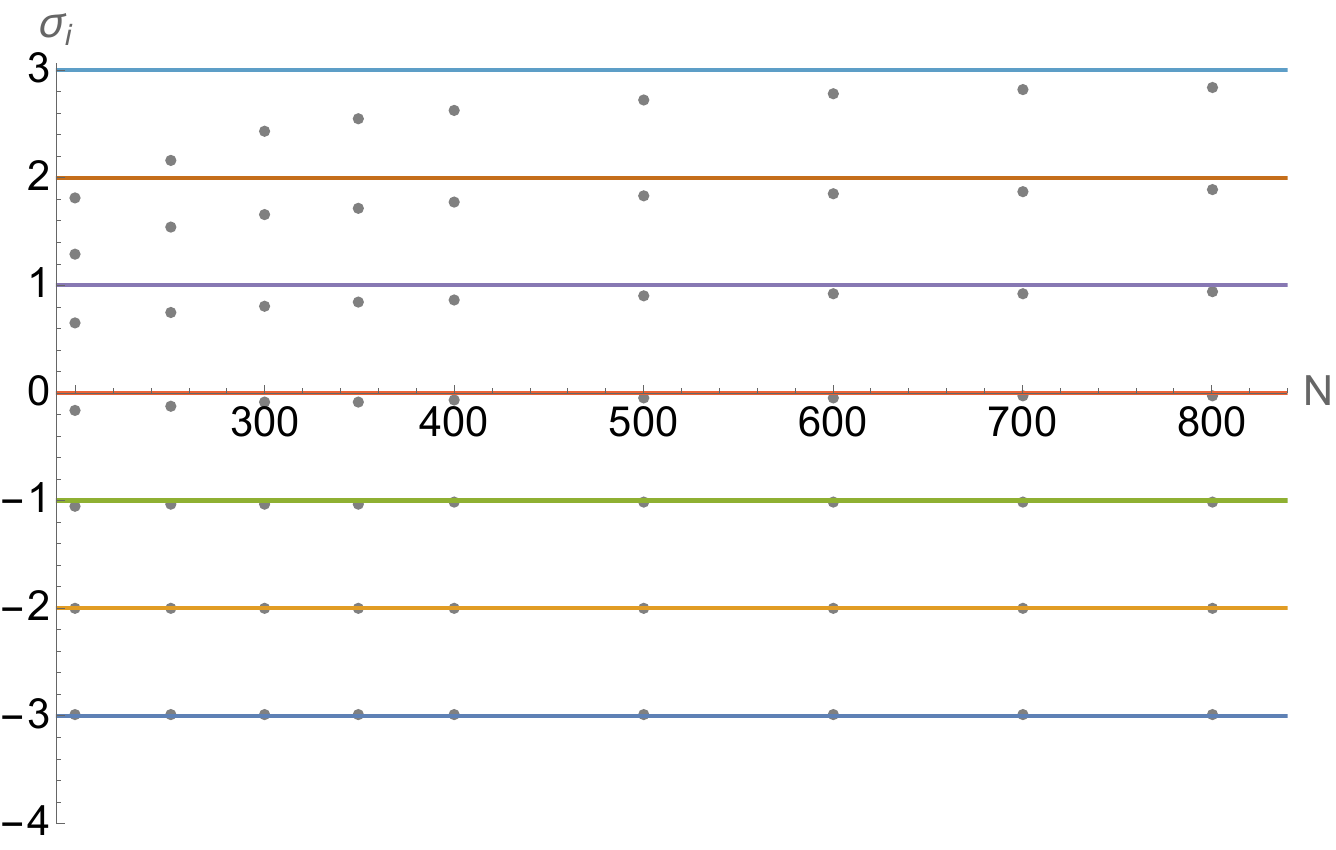}\caption{The seven most relevant eigenvalues between $-3$ and 3 of the linearized RG flow around the FP on the regular branch ${\cal A}$ followed as a function of $N$ along the hyperbola $N=\alpha/(3-d)$ with $\alpha=0.58$. This FP tends to a regular FP on the BMB line when $N\to\infty$ and the convergence of its eigenvalues to their expected values at $N=\infty$, that is, $3,2,1,0,-1,-2,-3$ is already clearly visible for $N\sim 800$.}
\label{fig:regular spectra}
\end{figure}

A detailed analysis of the eigenvalues and eigenfunctions associated with
singular FPs at $N=\infty$ was presented in Appendix~G of
Ref.~\citep{Yabunaka2022}.  However, because that study is both general and
technically involved, and does not directly address the fixed points on the
BMB line, we provide below a simplified derivation of the eigenvalue spectrum and eigenperturbations for the FPs of the tricritical BMB line at $N=\infty$ in $d=3$.

Our first goal is to understand how the critical exponent $\nu$ for
the FPs on the BMB line at $N=\infty$ can change discontinuously at
$\lambda=\lambda_{\rm BMB}$: specifically, $\nu = 1/2$ for
$0 \le \lambda < \lambda_{\rm BMB}$ and $\nu = 1/3$ at
$\lambda=\lambda_{\rm BMB}$. To determine $\nu$, we examine the linearized RG flow around the fixed points on both the regular and singular portions of the BMB line.

 Note that it is not easy to study the linearized flow around  singular FPs because  we do not know in general how to characterize the functional space of admissible eigenperturbations, see Appendix G of \citep{Yabunaka2022}. A way to overcome this difficulty is to regularize the singularities of the FPs, in which case we are back to the usual situation, and to continuously approach the singular FPs. A convenient way to achieve this goal is to study the FPs at finite (and large) $N$ that become the FPs of the  singular part of the BMB line when $N$ is sent to infinity. As long as $N$ is finite they are regular and the critical exponents attached to these FPs can be obtained from the usual linearization of the flow. The singularities of the FPs and of the eigenperturbations build up as $N\to\infty$ while the eigenvalues of the linearized flow vary smoothly with $N$ and have  each a well-defined limit when $N\to\infty$.

\begin{figure}
\includegraphics[scale=0.5]{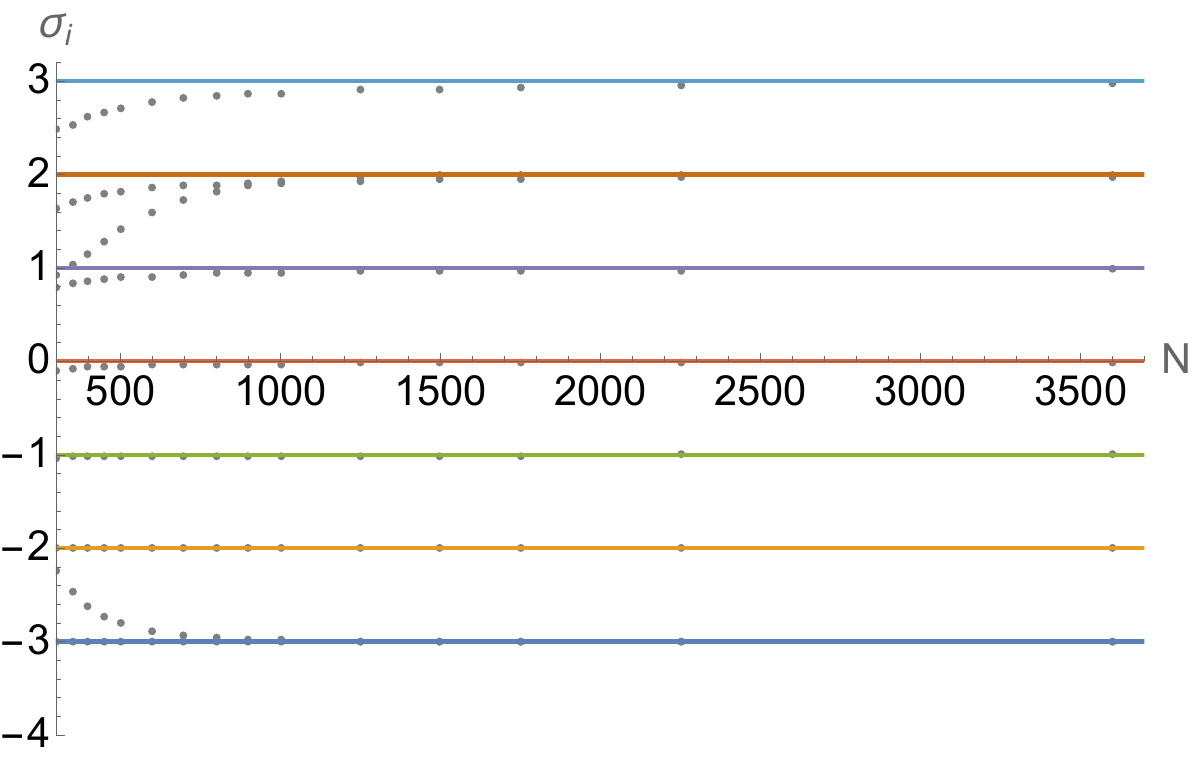}
\caption{The nine most relevant eigenvalues of the linearized RG flow around the FP on the singular branch $S{\cal A}$  followed as a function of $N$ along the hyperbola $N=\alpha/(3-d)$ with $\alpha=0.522$. These eigenvalues converge to their expected values, that is, $3,2,2,1,0,-1,-2,-3,-3$ when $N\to\infty$.} 
\label{fig: singular-eigenvalues}
\end{figure}

Before turning to the BMB fixed point (FP), we first verify how the eigenvalue spectrum of the regular tricritical FPs along the BMB line $(0 \leq \lambda < \lambda_{\rm BMB})$ is recovered from the finite-$N$ analysis outlined above. As reviewed in Sec.~\ref{finite-N-extension}, the FPs on the BMB line admit a finite-$N$ continuation when followed in the $(d,N)$ plane along hyperbolae parameterized by $\alpha = (3-d)N$. We work in the Wilson--Polchinski formulation of the renormalization group, which is particularly convenient for this purpose, and track one of the ${\cal A}$ 
 FPs corresponding to a fixed value of $\alpha$. We then compute the first seven eigenvalues associated with this FP.

Figure~\ref{fig:regular spectra} shows that these eigenvalues $\sigma_{i=1,2,\ldots}$ evolve smoothly with increasing $N$ and rapidly converge toward their expected large-$N$ limits, already closely approaching them for $N \sim 800$. In the $N \to \infty$ limit, the spectrum around the FP potential $\bar{V}\left(\bar{\varrho}\right)$ is given by \citep{litim2018asymptotic,litim2017fixed}
\begin{equation}
\left\{ \sigma_{i=1,2,\ldots} \right\}
= \left\{ -3, -2, -1, 0, 1, 2, 3, \ldots \right\},
\label{eq:regular spectra}
\end{equation}
and the eigenperturbations are given by \begin{equation}
\delta\bar{V}\left(\bar{\varrho}\right)\propto\left(\frac{\bar{V}'\left(\bar{\varrho}\right)}{1-\bar{V}'\left(\bar{\varrho}\right)}\right)^{\left(d+\sigma_{i}\right)/2}.
\end{equation}


Note that, in our convention, the most relevant eigendirections correspond to negative eigenvalues with the largest absolute values. The eigenvalue $-3$ is trivial: it is associated with the eigenperturbation $\delta \bar V(\bar\varrho)=\mathrm{const}$ and simply equals $-d$ with $d=3$. The most relevant nontrivial eigenvalue is $-2$, which determines the critical exponent $\nu=1/2$. 
\footnote{At finite $N$, one eigenvalue lies slightly below zero and converges to zero as $N\to\infty$. This explains how the $\tilde A$  FP can possess three relevant eigendirections at finite $N$, while it converges in the large-$N$ limit to a FP on the tricritical line with two relevant eigendirections and one marginal direction. For the $A$ FP, the eigenvalue slightly above zero converges to zero from above as $N\to\infty$, which explains why $A$ has only two relevant directions  at finite $N$.}
The spectrum of eigenvalues reported in Eq.~\eqref{eq:regular spectra} for $\alpha=0.58$ is in fact independent of $\alpha$, provided the FP is regular.

\begin{figure}
\includegraphics[scale=0.7]{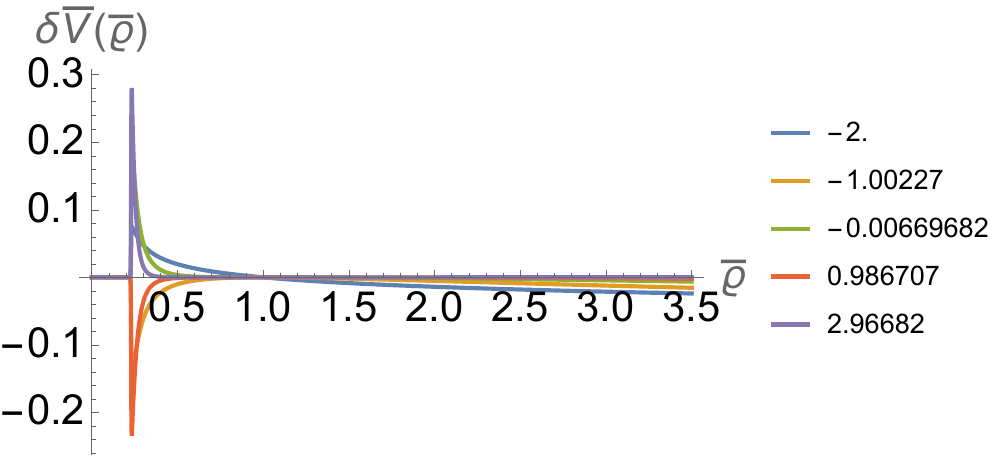} \caption{Eigenpertubations corresponding to the non degenerate eigenvalues  of the linearized Polchinski RG flow around the FP on the singular branch $S{\cal A}$ 
for $\alpha=0.522$ at $N=3600.$ The different colors correspond to the eigenvalues of the series given in Eq.~\eqref{eigenvalues-singular}. When $N\to\infty$, these eigenperturbations are vanishing for $\bar{\varrho}<\bar{\varrho}_{0}$ and are identical to the eigenperturbations of the regular FPs for $\bar{\varrho}>\bar{\varrho}_{0}$. }
\label{fig:singular-eigenperturbations}
\end{figure}

We now turn to the eigenvalues associated with the BMB fixed point. This FP is difficult to access directly because the numerical convergence to the large-$N$ limit becomes increasingly slow as one approaches the BMB point, and is most likely infinitely slow exactly at the BMB point itself. We therefore access it indirectly by approaching it along the singular branch of the BMB line by choosing $\alpha = 0.522$, slightly above $\alpha_{\rm BMB} \simeq 0.516$. For $N = 3600$, this corresponds to $d = 3 - \alpha/N = 2.999855$, and the nine most relevant 
\begin{equation}
\label{eigenvalues-singular}
\begin{split}
\left\{ \sigma_{i=1,2,\ldots} \right\}
= \{ & -2.999855, -2.999855, -2.00000, -1.00227, \\
& -0.00669683, 0.986707, 1.9779, 2.00000, 2.96682 \}.
\end{split}
\end{equation}
We observe that the first two eigenvalues are numerically degenerate, within the limits of numerical precision, for $N = 3600$.
As in the previous cases, these eigenvalues are already very close to their asymptotic values, see Fig.~\ref{fig: singular-eigenvalues}, given by:
\begin{equation}
\left\{ \sigma_{i=1,2,\ldots} \right\}
= \left\{ -3, -3, -2, -1, 0, 1, 2, 2, 3, \ldots \right\}.
\label{eq:singular spectra}
\end{equation}
This spectrum of eigenvalues clearly differs from the spectrum on the regular BMB line, Eq.~\eqref{eq:regular spectra}, due to the double degeneracy of the eigenvalues $-3$ and $2$.
However, here again, provided the FP is singular, the eigenvalue spectrum becomes at $N=\infty$ independent of the value of $\alpha$. We have also checked that the corresponding eigenperturbations exhibit similarly good convergence: their shapes become stable as $N$ increases, in full analogy with the regular  FPs discussed above. 

\begin{figure}
\includegraphics[scale=0.6]{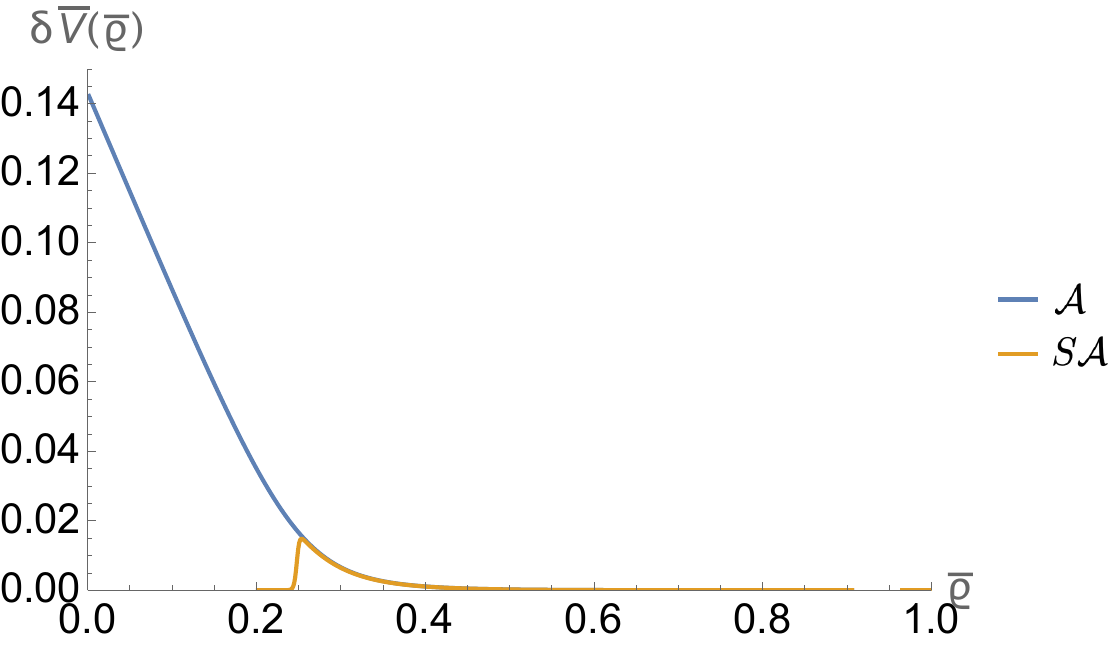}\caption{Eigenpertubations corresponding to the  eigenvalue
0.965559 of the linearized Polchinski RG flow around   the two FPs on the regular branch ${\cal A}$ and singular branch $S{\cal A}$ for the same value of $\alpha=0.58$ at $N=1400.$ The value of $\bar\varrho_0$ is $0.246$ for $\alpha=0.58$ at $N=\infty$.} 
\label{fig:eigenA3SA4}
\end{figure}

\begin{figure}
\includegraphics[scale=0.5]{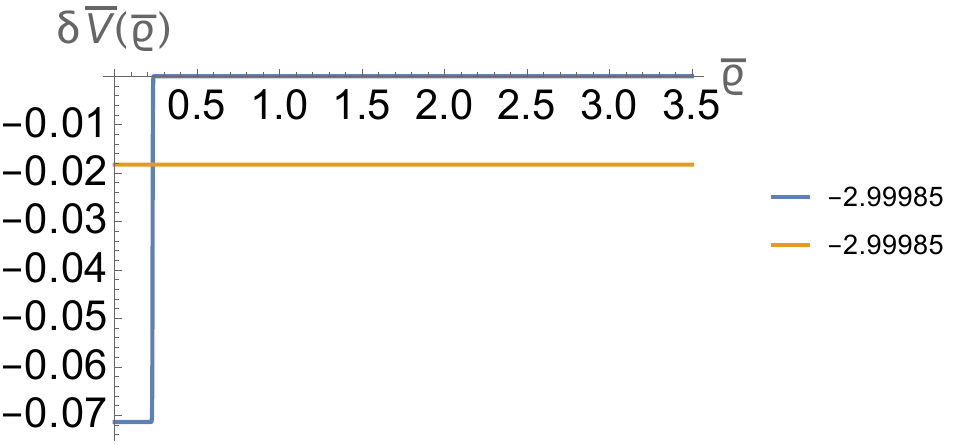}
\includegraphics[scale=0.5]{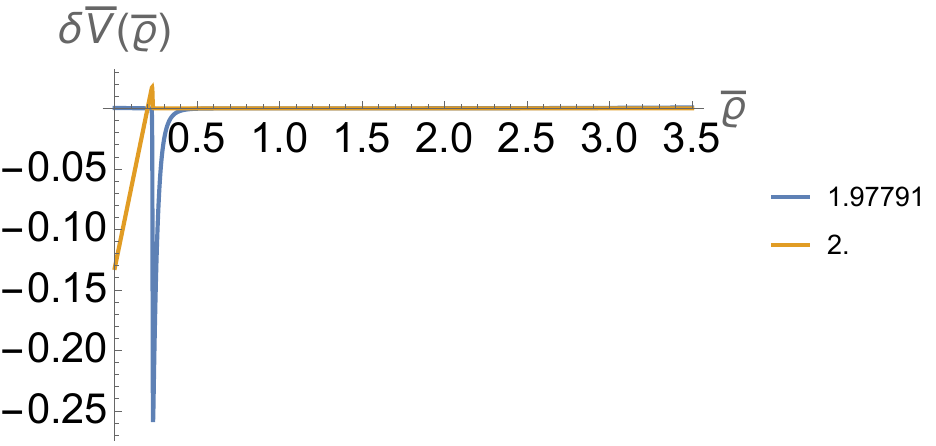}
\caption{ Eigenpertubations  around the FP on the singular branch $S{\cal A}$  for $\alpha=0.522$  and $N=3600$   of the linearized Polchinski RG flow. (Left) Eigenperturbations corresponding to the two numerically degenerate eigenvalues -2.999855. (Right) Eigenpertubations corresponding to the almost degenerate eigenvalue
1.97791 and 2.}
\label{fig:eigen3}
\end{figure}
In Fig.~\ref{fig:singular-eigenperturbations}, we display the subset of eigenfunctions at finite $N$ corresponding to the eigenvalues that remain nondegenerate in the $N \to \infty$ limit, namely $\{-2, -1, 0, 1, 3\}$. These eigenvalues also belong to the spectrum of regular FPs, see Eq.~\eqref{eq:regular spectra}.  As $N$ increases, the eigenperturbations associated with these eigenvalues converge rapidly; that is, they become practically independent of $N$, and a true discontinuity appears for each of them as $N\to\infty$. In this limit, the eigenperturbations are constant for $\bar\varrho < \bar\varrho_{0}$ and vanish for $\bar\varrho > \bar\varrho_{0}$.

A direct comparison of the eigenfunctions associated with  these eigenvalues for both the singular and the regular FPs, taken at the same values of $N$ and $\alpha$, shows that they coincide for $\bar\varrho > \bar\varrho_{0}$, see Fig.~\ref{fig:eigenA3SA4}  where $\bar\varrho_{0}=0.229$ for $\alpha=0.522$. The distinction appears for $\bar\varrho < \bar\varrho_{0}$: for singular FPs, the eigenfunctions  decrease very rapidly and almost vanish for $\bar\varrho < \bar\varrho_{0}$, whereas for regular FPs they remain smooth and nonvanishing throughout this region.

As for the two numerically degenerate eigenvalues $-2.999855$, one is associated with the trivial eigenperturbation
$\delta \bar V(\bar\varrho) \simeq {\rm const}$ and the other one corresponds to a nontrivial eigenperturbation, shown in Fig.~\ref{fig:eigen3}.  At $N=\infty$, this eigenperturbation is constant for $\bar\varrho < \bar\varrho_{0}$ and vanishes for $\bar\varrho > \bar\varrho_{0}$. Since for FPs on the singular part of the BMB line it is the most relevant among the nontrivial eigenvalues, it controls the correlation-length exponent, yielding $\nu = 1/3$ in $d=3$ at $N = \infty$.

As for the two eigenvalues 2, we show their associated eigenfunctions in Fig.~\ref{fig:eigen3}. As $N$ increases the shape of these functions stabilizes and a genuine singularity appears at $N=\infty$. Here again, a direct comparison of the eigenfunctions associated with this eigenvalue  for the singular and regular FPs, taken at the same values of $N$ and $\alpha$, shows that they coincide for $\bar\varrho > \bar\varrho_{0}$ while they identically vanish for $\bar\varrho < \bar\varrho_{0}$ for singular FPs and remain continuous and nonvanishing throughout this region for regular FPs.

For the part of the spectrum made of twice degenerate eigenvalues, that is $\left\{ -3,2,\cdots\right\} $, we have already shown that for each eigenvalue, one of the eigenperturbation is non-vanishing  for $\bar\varrho\in[0,\bar\varrho_0]$ only, see Figs.~\ref{fig:eigen3}. On this interval, the FP potential is linear, see Fig.~\ref{sing-pot-polchinski}. Therefore, it is not surprising that on this interval of $\bar\varrho$, these eigenperturbations are identical to  the eigenperturbations of the linear FP: $\bar{V}\left(\bar{\varrho}\right)=\bar{\varrho}$:
\begin{equation}
C\left(\bar{\varrho}-\frac{1}{5}\right)^{n},\ \ \ n=0,1,2\cdots
\label{eq:eigen_pert_analytical}
\end{equation}
 whose  eigenvalues are, see Appendix G of \citep{Yabunaka2022}:
\begin{equation}
\label{eq:eigenvalues-linear}
\lambda_{n}=-3+5n.
\end{equation}
The above arguments show why the singular FPs in the Polchinski version of the RG has more eigenvalues than the regular FPs: the spectrum of eigenvalues is the union of the spectra of eigenvalues associated with the regular FPs and of the linear FP. Although we have only studied the first two degenerate eigenvalues of this series, that is $-3$ and 2, this holds for all the other values of $n$ in the series of eigenvalues given by Eq.~\eqref{eq:eigenvalues-linear}. 

A subtlety arises from the fact that the BMB FP is located at the junction of the two branches ${\cal {A}}$ and $S{\cal {A}}$ of the BMB line of FPs. 
We have shown that, except at the BMB FP, the correlation-length exponent is $\nu = 1/3$ along the entire $S{\cal {A}}$ branch, whereas $\nu = 1/2$ for all fixed points on the ${\cal {A}}$ branch. 
At $N=\infty$ and $d=3$, the BMB FP can be approached from either branch, ${\cal {A}}$ or $S{\cal {A}}$. 
Alternatively, one may define the BMB FP as the large-$N$ limit of a finite-$N$ fixed point, either  ${\cal {A}}$ or $S{\cal {A}}$.  This suggests that the critical exponent $\nu$ at the BMB FP itself is not uniquely determined, but rather depends on the path by which the limit $N \to \infty$ and $d \to 3$ is taken. 

\section{Flow of the dimensionful mass\label{sec:Flow-of-dimensionfull}}

The physical dimensionful mass, that is, the inverse correlation length, is the square root of the curvature of the effective potential at the origin. This means that this mass is the limit when $k\to 0$ of the running mass $m_k$ given by:
\begin{equation}
   m^2=\lim_{k\to 0} m_k^2=\lim_{k\to 0} U_k''(\phi)_{\vert\phi=0} =\lim_{k\to 0} k^2 \tilde{U}_{k}''(\tilde{\phi})_{\vert\tilde\phi=0}=\lim_{k\to 0} k^2 \bar{U}_{t}''(\bar{\phi})_{\vert\bar\phi=0}
\end{equation}
where we have ignored the field renormalization $\bar Z_k$ because we are working with the LPA.

In this section, we study the flow of the dimensionful mass using the effective potential in both formulations: $\bar{U}_{t}(\bar{\phi})$ in the Wetterich parametrization and $\bar{V}_{t}(\bar{\varrho})$ in the Wilson-Polchinski parametrization.

As long as $t$ is finite, the potentials $\bar{U}_{t}$ and $\bar{V}_{t}$  are regular and can be expanded as a power series of respectively $\bar{\phi}^2$ and $\bar{\varrho}$:
\begin{equation}
\label{eq: expansion-U}
\bar{U}_{t}\left(\bar{\phi}\right)=\bar{U}_{t}\left(\bar{\phi}=0\right)+\frac{\tilde{a}_{t}^{W}}{2}\bar{\phi}^{2}+\cdots
\end{equation}
\begin{equation}
\bar{V}_{t}\left(\bar{\varrho}\right)=\bar{V}_{t}\left(\bar{\varrho}=0\right)+\tilde{a}_{t}^{P}\bar{\varrho}+\cdots
\end{equation}
and so

\begin{equation}
m_k^2=k^{2}\tilde{a}_{t}^{W}.
\end{equation}
The relation between $\tilde{a}_{t}^{W}$ and $\tilde{a}_{t}^{P}$
can be obtained from Eq. (\ref{eq:transformation-WP-W}):
\begin{equation}
\tilde{a}_{t}^{W}=\frac{\tilde{a}_{t}^{P}}{1-\tilde{a}_{t}^{P}},
\end{equation}
 which clearly
shows that  $\tilde{a}_{t}^{P}$ must be smaller than 1 and that the limit $\tilde{a}_{t}^{P}\to 1$ corresponds to $\tilde{a}_{t}^{W}\to\infty$.

\begin{figure}
\includegraphics[scale=0.4]{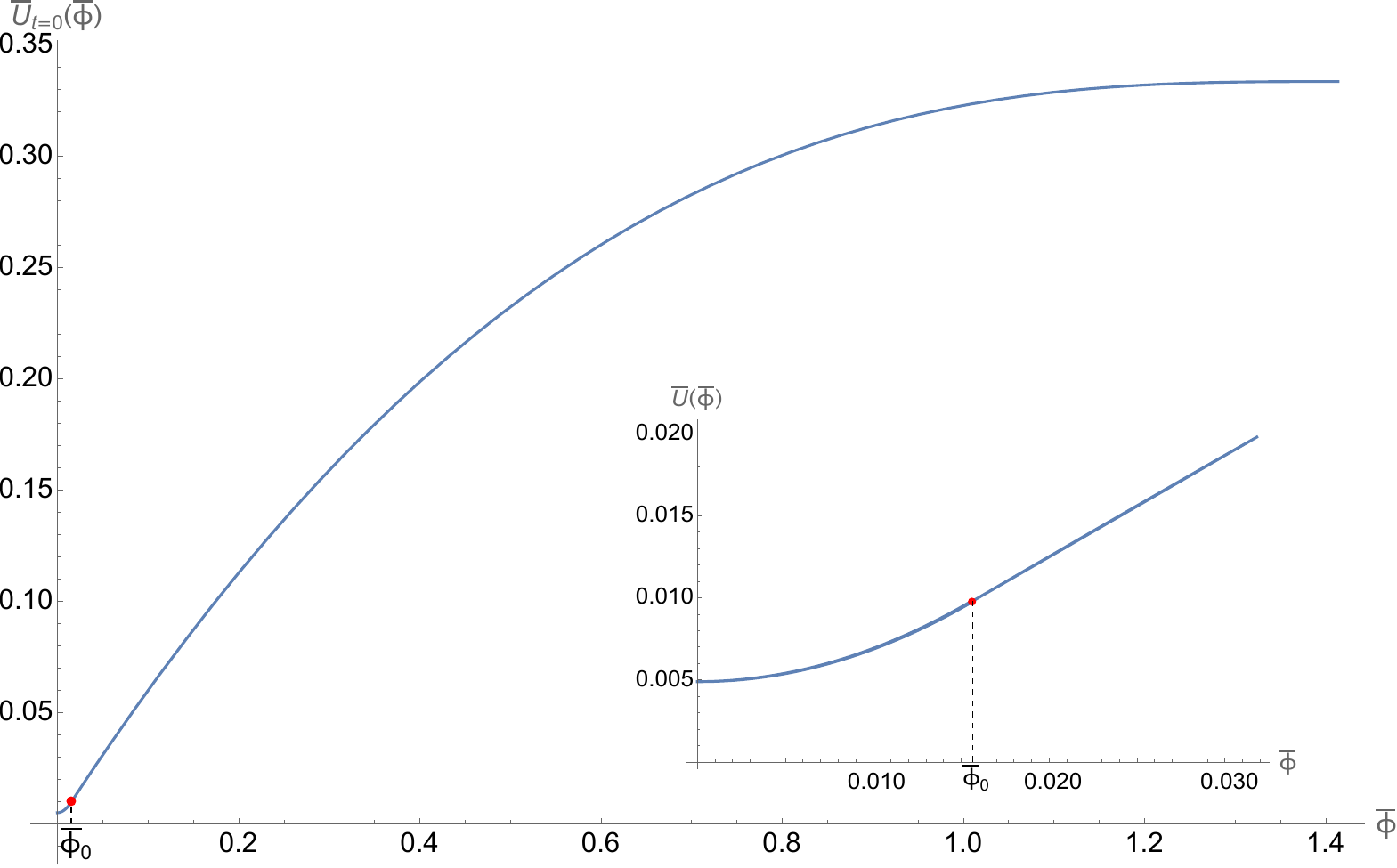}\caption{Initial condition $\bar{U}_{t=0}\left(\bar{\phi}\right)=\bar{U}_{BMB}\left(\bar{\phi}\right)+\delta\bar{U}\left(\bar{\phi}\right)$ with $C=1/41$, where $C$ is defined in Eq. (\ref{eq:deltaV}). Under the transformation given by Eqs. (\ref{eq:transformation-WP-W}) and (\ref{eq:transform-WP-W-2}), $\delta\bar{U}\left(\bar{\phi}\right)$ is the eigenperturbation corresponding to the eigenvalue 2 in the spectrum given in Eq.~(\ref{eq:eigenvalues-linear}).  } %
\label{fig:Ut=0}
\end{figure}

By perturbing the BMB FP along its least irrelevant singular eigendirection, we investigate the mechanism of mass generation at the BMB FP. This eigenperturbation corresponds to the eigenvalue $2$ in the spectrum given in Eq.~(\ref{eq:eigenvalues-linear}). In the Wilson--Polchinski formulation, it is linear and reads
\begin{equation}
\delta\bar{V}(\bar{\varrho})=-C\left(\bar{\varrho}-\tfrac{1}{5}\right).
\label{eq:deltaV}
\end{equation}
We choose $0<C\ll1$ so that $\delta\bar{V}(\bar{\varrho})$ represents an infinitesimal perturbation, allowing the RG flow to be linearized in the vicinity of the BMB FP. The initial condition for the flow is therefore $\bar{V}_{t=0}(\bar{\varrho})=\bar{V}_{\mathrm{BMB}}(\bar{\varrho})+\delta\bar{V}(\bar{\varrho})$. Since both $\delta\bar{V}(\bar{\varrho})$ and $\bar{V}_{\mathrm{BMB}}(\bar{\varrho})$  are linear for $0\leq\bar\varrho\lesssim\bar\varrho_{0}$, see Fig. \ref{fig:FP potentials on the BMB line WP}, so is $\bar{V}_{t=0}(\bar{\varrho})$ and the slope is $1-C$. This linear region is mapped onto a very narrow region of the initial potential in the Wetterich formulation: $\bar{U}_{t=0}\left(\bar{\phi}\right)=\bar{U}_{BMB}\left(\bar{\phi}\right)+\delta\bar{U}\left(\bar{\phi}\right)$ for $0\leq\bar{\phi}<\bar{\phi}_{0}$,
where $\bar{\phi}_{0}\ll 1$, see Fig. \ref{fig:Ut=0}.

The slope $\bar{V}'_{t}\left(\bar{\varrho}=0\right)$ at the origin  satisfies 
\begin{equation}
\label{eq:slope-origin-V}
\tilde{a}_{t}^{P}=\bar{V}'_{t}\left(\bar{\varrho}=0\right)=1-Ce^{2t}
\end{equation}
since $C$ is very small and $\delta V$ is an eigenperturbation.
Translated in the Wetterich parametrization, we find from Eqs. \eqref{eq: expansion-U} - \eqref{eq:slope-origin-V}
\begin{equation}
m^2_t=\Lambda^2 e^{2t} \frac{1-Ce^{2t}}{Ce^{2t}} \stackrel[t\to-\infty]{}{\longrightarrow} C^{-1}\varLambda^{2},
\end{equation}
which shows that even though the flow runs towards a fixed point --- the BMB FP ---  the physical mass at this fixed point is finite and large. We now show how to extend the previous result to arbitrary masses.

To address this question, we analyze the flow of the potential starting from an initial condition $\bar{V}_{t=0}(\bar{\varrho})$ for which the slope at the origin, i.e., $\bar{V}'_{t=0}(\bar{\varrho}=0)$, deviates from unity by a finite amount.
We limit ourselves to the BMB FP because the analytical solution at
Large-$N$ is only available around this FP and not around the
other singular FPs. Hereafter, we only use the Wetterich parametrization
and the Large-$N$ flow of $\bar{U}_{t}\left(\bar{\phi}\right)$, Eq. (\ref{eq:LargeNflow}).

When the initial condition $\bar{U}_{t=0}\left(\bar{\phi}\right)$
 satisfies $\bar{U}'_{t=0}\left(\bar{\phi}\right)>0$
for all $\bar{\phi}>0$, the flow of $\bar{U}_{t}\left(\bar{\phi}\right)$
is given by the following implicit expression \citep{Mati2017}:
\begin{equation}
\bar\rho=1+\sqrt{\bar{U}_{t}'\left(\bar{\rho}\right)}\left(F\left(\bar{U}_{t}'\left(\bar{\rho}\right)\right)+G\left(\bar{U}_{t}'\left(\bar{\rho}\right)e^{2t}\right)\right)\label{eq:flow of pot}
\end{equation}
where $F(x)$  is given by Eq. (\ref{eq:Fu'}). 
The function $G\left(x\right)$ is fixed by the initial condition
\begin{equation}
G\left(x\right)=\frac{\bar\rho_{\varLambda}\left(x\right)-1}{\sqrt{x}}-F\left(x\right)\label{eq:G(x)}
\end{equation}
where $\bar\rho_{\varLambda}\left(x\right)$ is the inverse function of
$\bar{U}_{t=0}'\left(\bar{\rho}\right)$ and satisfies $\bar\rho_{\varLambda}\left(\bar{U}_{t=0}'\left(\bar{\rho}\right)\right)=\bar\rho$. 

We now discuss a necessary  condition for an initial condition $\bar{U}_{t=0}(\bar\rho)$ to reach the BMB FP at long RG time. 

We first consider a field expansion of the potential around $\bar{\rho}=1$ because the flow of the coupling constants of this expansion shows a very nice property specific to  $N=\infty$. We work with the derivative of the potential and expand it at order $n$:
\begin{equation}
   \bar{U}_{t}'\left(\bar{\rho}\right)=\left(\bar\rho-1\right)^{2}/c_t^{2}+a_{3,t}\left(\bar\rho-1\right)^{3}+ \cdots +a_{n,t}\left(\bar\rho-1\right)^{n}+ O\left(\left(\bar\rho-1\right)^{n+1}\right).
   \label{potential-expanded} 
\end{equation}
It is easy to show using Eq.~\eqref{eq:flow-LPA} that whatever the value of $n$, the flow of the couplings $(c_t,\cdots, a_{p,t})$ with $p\le n$ does not depend on the couplings $a_{p',t}$ with $p'>p$.  It is also easy to show that provided $\bar{U}_{t}'$ involves neither a constant term nor a term linear in $\bar\rho-1$, as in Eq.~\eqref{potential-expanded}, the flow of $c_t$ vanishes and all other couplings flow to their fixed point value when $t\to-\infty$. This means that any initial condition such that:
\begin{equation}
\bar{U}_{t=0}'\left(\bar{\rho}\right)=\left(\bar\rho-1\right)^{2}/c^{2}+O\left(\left(\bar\rho-1\right)^{3}\right),\label{eq:initial_expansion}
\end{equation} 
evolves to a FP along the BMB line indexed by the parameter $c=c_{t=0}$.

It is important to notice here that although exact the condition given in Eq.~\eqref{eq:initial_expansion} is only necessary but is not sufficient because the Taylor expansion of $\bar{U}_{t}'\left(\bar{\rho}\right)$ around $\bar\rho=1$ has a finite radius of convergence and we need to have a global solution, valid for all field values, to prove that the dimensionful mass at the BMB FP is nonuniversal and can be tuned at will. Moreover, we know that the BMB line has a finite extension, with $c<c_{\rm BMB}$, a feature that is not seen in the the argument above and that requires to have a global solution for the FP potential rather than a Taylor expansion.

\begin{figure}
\includegraphics[scale=0.5]{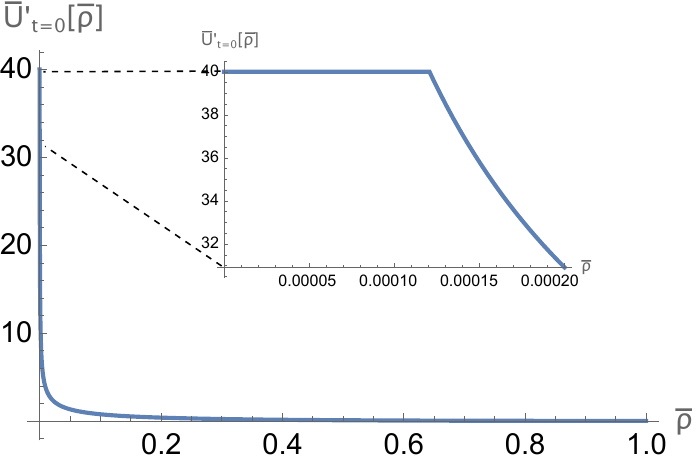}\includegraphics[scale=0.5]{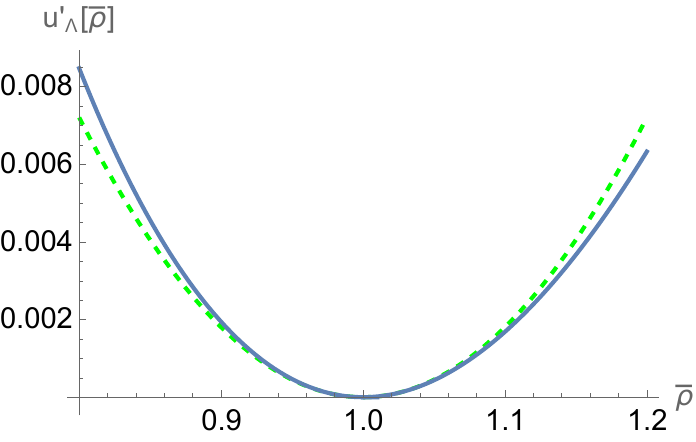}\caption{Initial condition  $\bar{U}_{t=0}'\left(\bar{\rho}\right)$ of Eq.~\eqref{eq:initialcondition} with $M_0^2 \Lambda^{-2}=40$ and $\varepsilon=10^{-3}$. (Left) Plot for $0\le\bar{\rho}\le1$. (Right) Plot around $\bar{\rho}=1$. The green dashed curve corresponds to $(\bar{\rho}-1)^2/c_{\rm BMB}^2$. }  
\label{fig:initial_condition}
\end{figure}

\begin{figure}
\includegraphics[scale=0.5]{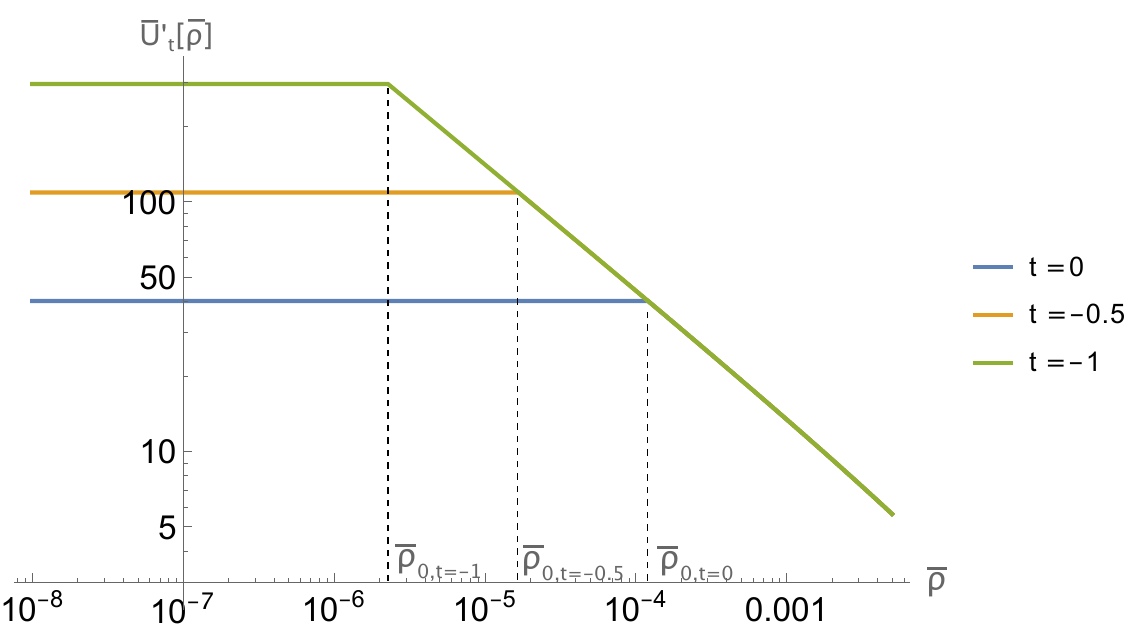}
\caption{Plot of $\bar{U}_{t}'\left(\bar{\rho}\right)$ at different times. The initial condition is given by Eq.~\eqref{eq:initialcondition} with $M_0^2 \Lambda^{-2}=40$ and $\varepsilon=10^{-3}$.  For $\bar\rho<\bar\rho_{0,t}$, $\bar{U}_{t}'\left(\bar{\rho}\right)$ is almost constant and equal to $M_0^2 \Lambda^{-2} \exp(-2t) $ while  it coincides with $\bar U'_{BMB}(\bar\rho) $ for $\bar\rho>\bar\rho_{0,t}$. The divergence of $\bar{U}_{t}'\left(\bar{\rho}\right)$ at $\bar\rho=0$ builds up in the limit $k\to 0$. }  
\label{fig:dimlesspot_evolution}
\end{figure}

\begin{figure}
\includegraphics[scale=1]{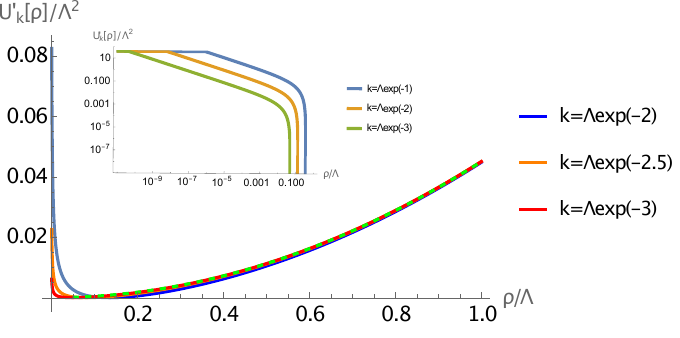}

\includegraphics[scale=0.7]{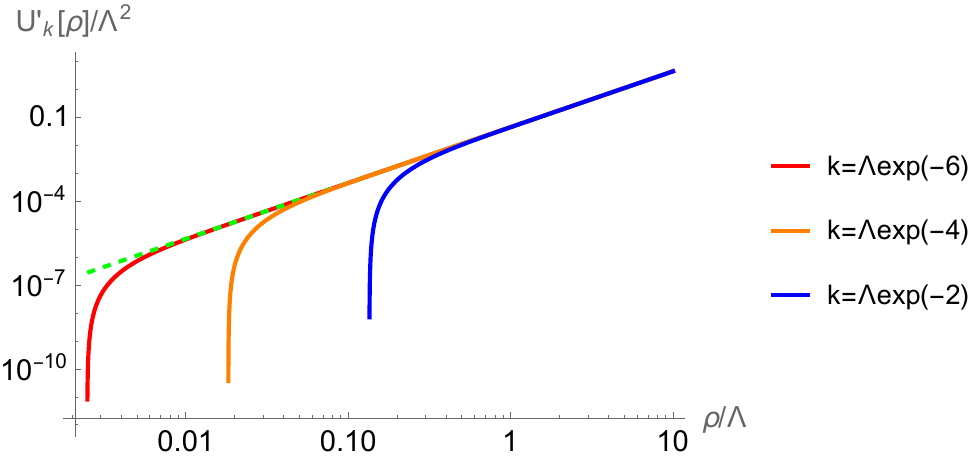}
\caption{Evolution of  the derivative of the dimensionful running potential $U_k'(\rho)$ with $M_0^2 \Lambda^{-2}=40$ and $\varepsilon=10^{-3}$.  (Top) Behavior around $\rho=0$. (Bottom) Approach of $U_k'(\rho)$ towards  the BMB potential $U'(\rho)=\rho^2/(4 c_{\rm BMB}^2)$,  valid for $\rho$ strictly positive and shown as the green dashed curve. This limiting behavior is realized in the limit $k\to 0$ for each $\rho>0$, while $U_k'(\rho=0)$ remains almost $M_0^2$ all along the flow. } 
\label{fig:dimfulpot_evolution}
\end{figure}

\begin{figure}
\includegraphics[scale=0.65]{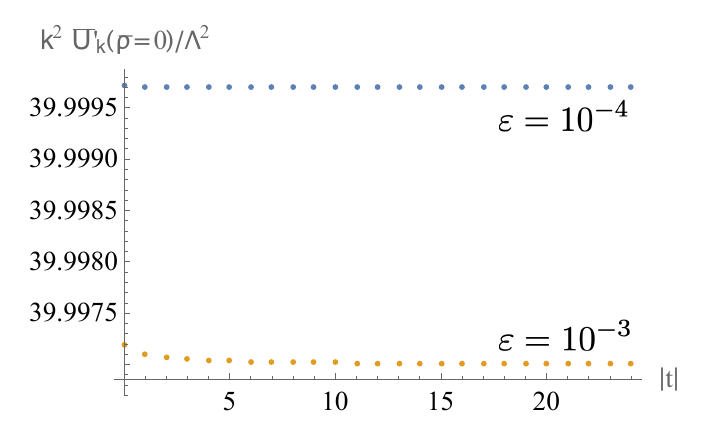}\includegraphics[scale=0.5]{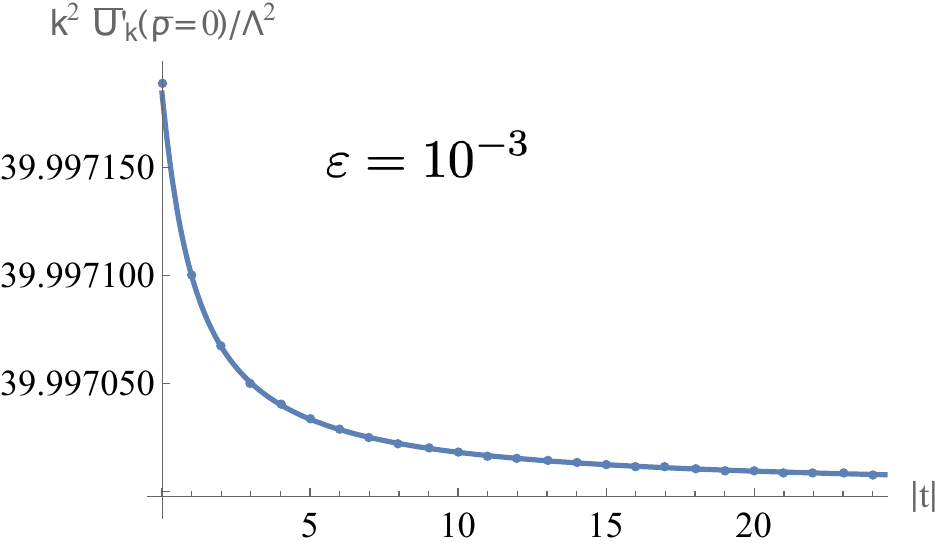}\caption{(Left) Evolution of the square of the dimensionful mass divided by $\Lambda^2$ as a function of the renormalization time
$t$ for $\varepsilon=10^{-4}$ and $10^{-3}$. 
 As $\varepsilon$ decreases, the flow becomes slower and slower and eventually vanishes in the limit $\varepsilon\to 0$. (Right) Comparison between the square of the dimensionful mass divided by $\Lambda^2$ (indicated with points) and the RHS of Eq. \eqref{eq:asymptotic_t} (shown as a solid curve) for $\varepsilon=10^{-3}$.} 
\label{fig: evolution of mass}
\end{figure}
Thus, we look for a family of non-singular  initial conditions that reach the BMB FP and realize finite values of the physical dimensionful mass at $k=0$. We  use the
following initial condition:
\begin{equation}
\bar{\rho}_{\Lambda}=1+\sqrt{\bar{U}_{t=0}'\left(\bar{\rho}_{\Lambda}\right)}F(\bar{U}_{t=0}'\left(\bar{\rho}_{\Lambda}\right))-\frac{1}{4}(3\pi)\sqrt{\bar{U}_{t=0}'\left(\bar{\rho}_{\Lambda}\right)} +{\cal T}((-\bar{U}_{t=0}'\left(\bar{\rho}_{\Lambda}\right)+M_{0}^2\Lambda^{-2})/\varepsilon),
\label{eq:initialcondition}
\end{equation} 
which satisfies Eq.~\eqref{eq:initial_expansion} with $c=c_{\rm BMB}$. 
The function $\mathcal{T}(x)$ is defined as:
\begin{equation}
\mathcal{T}(x) = 
\begin{cases} 
\tanh(x) - 1 & (x \le x_1) \\
(\tanh(x) - 1) \frac{\psi(x_2-x)}{\psi(x-x_1) + \psi(x_2-x)} & (x_1 < x < x_2) \\
0 & (x \ge x_2)
\end{cases}
\label{eq:smooth_cutoff}
\end{equation}
 with
$\psi(x)$ a standard mollifier satisfying $\psi(x) = \exp(-1/x)$ for $x > 0$ and $\psi(x) = 0$ for $x \le 0$. Taking $x_1=2$ and $x_2=3$, we find that $\bar{U}'_{t=0}$  coincides with $\bar{U}'_{\rm BMB}$ for $\bar{U}'_{\rm BMB}(\bar\rho)<M_{0}^2\Lambda^{-2}-3\varepsilon$ while the divergence  of this FP potential at the origin is cutoff when  $\bar{U}'_{\rm BMB}$ becomes larger than $M_{0}^2\Lambda^{-2}$, see the inset in Fig.~\ref{fig:initial_condition}. 

In Eq. (\ref{eq:initialcondition}), the initial condition is parameterized by $\varepsilon$ which represents the sharpness of the cutoff and $M_0$ which is very close to the bare mass for $\varepsilon\ll1$. The value of $\bar\rho$ where $\bar{U}'_{t=0}$ does no longer coincide with $\bar{U}'_{\rm BMB}$ is  given by  $\bar{U}_{t=0}'(\bar\rho_{0,t=0})=M_{0}^2\Lambda^{-2}-3\varepsilon$. For the parameters $M_{0}^2\Lambda^{-2}=40$ and $\varepsilon=10^{-3}$ used in Figs.~\ref{fig:initial_condition} and \ref{fig:dimlesspot_evolution}, $\bar{U}'_{t=0}$ coincides with the BMB FP for $\bar{\rho}>10^{-3}$ while the divergence of the BMB FP around $\bar{\rho}=0$ is cutoff. This makes the initial  dimensionful   mass at $t=0$ finite and very close to $M_{0}$.

The important point is that the value of $\bar\rho_{0,t}$ decreases all along the flow but remains finite for all $t$ --which makes $\bar{U}'_{t}$ approaching the BMB FP potential for all $\bar\rho> \bar\rho_{0,t}\to 0$ when $t\to-\infty$, see Fig.~\ref{fig:dimfulpot_evolution}-- while $\bar{U}'_{t}(\bar\rho=0)$ becomes larger and larger, see Fig.~\ref{fig:dimlesspot_evolution}, in such a way that the singularity at the origin builds up when $t\to-\infty$. We thus find that, in the limit $\varepsilon\to 0$, $\bar{U}_{t}'\left(\bar{\rho}\right)=M_0^2 \Lambda^{-2} \exp(-2t)$ for $\bar\rho< \bar\rho_{0,t}$ and $\bar{U}_{t}'\left(\bar{\rho}\right)=\bar{U}_{\rm BMB}'\left(\bar{\rho}\right)$ for $\bar\rho> \bar\rho_{0,t}$ which is an exact  solution of Eq.~\eqref{eq:LargeNflow} in the limit $\varepsilon\to 0$.
Simultaneously,  the dimensionful mass remains almost constant and, more precisely, it evolves even less as $\varepsilon$ becomes smaller, see Fig.~\ref{fig: evolution of mass}. As a result,  the critical surface associated with the BMB FP is approached as $\varepsilon\to0$   and the physical dimensionful mass, obtained when $k=0$, approaches $M_0$.

We can resolve analytically the flow of the square of the dimensionful mass when $t\to -\infty$ starting from the initial condition given by Eq.~\eqref{eq:initialcondition}. As shown below, we find:
\begin{equation}\label{eq:asymptotic_t}
U'_k(\rho=0)\Lambda^{-2} \approx M_0^2 \Lambda^{-2}-x_2 \varepsilon + \varepsilon \left[ \frac{1}{5|t |+ C_1} - \frac{K}{(5|t| + C_1)^3} \right]+ O(|t|^{-3})
\end{equation} 
 where $C_1=\ln(5M_0^4 \Lambda^{-4}) + 1 + \ln(1-\tanh x_2)$ and
$K = 1+\text{sech}^2 x_2 / (1-\tanh x_2) + 2 \varepsilon/(M_0^2 \Lambda^{-2})$. This means that the square of the dimensionful mass approaches $M_0^2-x_2 \varepsilon \Lambda^2$ algebraically when $t\to -\infty$.  

The derivation of Eq. \eqref{eq:asymptotic_t} goes as follows. We define $s = e^{t}$, $x = \bar{U}_t'(\bar{\rho} =0)$ and $y=\bar{U}_{t}'(\bar\rho =0) s^2$, where $t=\log(k/\Lambda)<0$ is the RG time. Using Eq. \eqref{eq:flow of pot}, $y$  is found to satisfy:
\begin{equation}
s + \sqrt{y} \left[ F(y/s^2) + \frac{{
\bar{\rho}_{\Lambda}}(y) - 1}{\sqrt{y}} - F(y) \right] = 0.
\label{eq:determining_y}
\end{equation}
and $F(x)$ behaves at large $x$ as 
\begin{equation}
F(x) = \frac{3\pi}{4} - x^{-1/2} + 0 \cdot x^{-3/2} + \frac{1}{5}x^{-5/2} + O(x^{-7/2}).
\end{equation}
Since $\bar{U}_t'(\bar{\rho})$ diverges for $\bar{\rho}=0$ when $t\to-\infty$ due to the singularity of the BMB FP as shown in Fig. \ref{fig:dimlesspot_evolution}, we can substitute $F(y/s^2)$ by its expansion  for large $y/s^2$. We thus obtain
\begin{equation}
 \frac{3\pi}{4}\sqrt{y}  + \frac{s^5}{5y^2} + \bar\rho_{\Lambda}(y) - 1 - \sqrt{y}F(y) = 0,
\end{equation}
where the leading order term in $s$ has canceled. By inserting in the previous equation the definition of $\bar\rho_{\Lambda}(y)$: $\bar\rho_{\Lambda}(y) = 1+ \sqrt{y}F(y) - \frac{3\pi}{4}\sqrt{y} + \mathcal{T}(z) $ with $z = -(y-M_0^2 \Lambda^{-2})/\varepsilon$, obtained from Eq. \eqref{eq:initialcondition}, we get:
\begin{equation}
\frac{s^5}{5y^2} + \mathcal{T}(z)  = 0 \implies   \mathcal{T}(z) = -\frac{e^{5t}}{5y^2}\label{eq:T_and_s}
\end{equation}
The function $\mathcal{T}(z)$ can also be obtained for   $z \nearrow x_2$ from Eq.~\eqref{eq:smooth_cutoff} for  $x_1 < x < x_2$ and $\varepsilon\ll 1$:
\begin{equation}
\mathcal{T}(z) = -S(1 - \tanh z)
\end{equation}
with $S$ given by
\begin{equation}
S = \frac{\exp(-1/\delta)}{\exp(-1/(1-\delta)) + \exp(-1/\delta)} \approx e^{(1+\delta)} \cdot e^{-1/\delta}
\end{equation}
with $\delta \equiv x_2 - z$, when $\delta \to 0$.
Taking the logarithm of Eq. \eqref{eq:T_and_s} yields:
\begin{equation}
-\frac{1}{\delta} + 1 +\delta+ \ln(1-\tanh (x_2-\delta)) = 5t - \ln(5 y^2).
\label{eq:determining_delta}
\end{equation}
By expanding the previous equation in powers of $\delta$ we obtain:
\begin{equation}
\frac{1}{\delta} = 5|t| + C_1 + K \delta + O(\delta^2)
\end{equation}
with $C_1$ and $K$  given below Eq. \eqref{eq:asymptotic_t} and with $y \approx M_0^2 \Lambda^{-2}$.
Solving for $\delta$ iteratively:
\begin{equation}
\delta \approx \frac{1}{5|t |+ C_1} - \frac{K}{(5|t| + C_1)^3}\label{eq:delta_approx}
\end{equation}  up to the order of $O(|t|^{-3})$. 
With $\varepsilon=1/1000$ and $x_2=3$, 
$K = 1+\text{sech}^2 3 / (1-\tanh 3) + 1/20000 \approx 2 + \tanh 3 \approx 2.995$ and
 $C_1=\ln(8000) + 1 + \ln(1-\tanh 3) \approx  4.678$. Using the definitions of $\delta$ and $z$ given above Eqs. \eqref{eq:T_and_s} and \eqref{eq:determining_delta}, we have $x_2-\delta=z = -(y-M_0^2 \Lambda^{-2})/\varepsilon$. Recalling $y=\bar{U}_{t}'(\bar\rho =0) s^2$, defined above Eq. \eqref{eq:determining_y}, this leads to $\bar{U}_{t}'(\bar\rho =0) s^2=M_0^2 \Lambda^{-2}+(\delta-x_2)\varepsilon$. Substituting Eq.  \eqref{eq:delta_approx} into this, we finally arrive at Eq. \eqref{eq:asymptotic_t}. As shown in Fig. \ref{fig: evolution of mass}, the RHS of Eq.~\eqref{eq:asymptotic_t} gives a very good approximation of the flow of the square of the dimensionful mass $U_k'(\rho=0)$.  

As a conclusion, any dimensionful physical mass $M_0$ can be obtained at the end of the flow by tuning the initial condition as in Eq. (\ref{eq:initialcondition}) in the limit $\varepsilon\to0$, where the divergence of  $\bar{U}'(\bar{\rho})$ of the BMB FP above $M_0^2 \Lambda^{-2}$ is cutoff   such that $\bar{U}'_{t=0}(\bar{\rho})$ becomes completely flat around $\bar\rho=0$. The divergence of the BMB FP builds up around $\bar\rho=0$  in the $t\to-\infty$ limit and the dimensionless potential approaches the BMB FP, while the runnning dimensionful mass defined at $\rho=0$ remains $M_0$ for all $t>-\infty$.

\section{Conclusion and discussion}

By means of the functional renormalization group, we have shown that mass generation at the BMB fixed point originates from the non-analytic, ``cuspy'' structure of the effective potential at $\phi = 0$. Importantly, the generated mass is non-universal: its value is fixed by the specific tuning of the ultraviolet (UV) initial conditions, i.e., by the choice of the bare potential.

Although the mechanism of mass generation is rooted in the cusp of the fixed-point potential, its actual value is governed by a singular perturbation associated with the eigenvalue $2$. This singular mode enables the  mass to acquire a finite value, even as the effective potential itself flows toward the BMB fixed point in the limit $k \to 0$ for all strictly positive values of $\phi$. Consequently, the mass behaves as an effectively free parameter, inherited from the UV initial conditions. This provides a clear explanation for the non-universal breaking of scale invariance, despite the RG flow being attracted to a fixed point.

The present analysis relies on the Local Potential Approximation (LPA), which is known to be exact for the flow of regular potentials, but only approximate in the presence of singular ones. This approximation neglects higher-order contributions in the derivative expansion and it would therefore be valuable to assess the robustness of our results upon inclusion of such terms. Nevertheless, we expect our conclusions to remain at least qualitatively unchanged.

It would also be important to extend our analysis to large but finite values of $N$, rather than restricting to the strict $N \to \infty$ limit. In this regime, a boundary layer is expected to play a central role: as $k \to 0$, the singularity at vanishing field, $\bar{\phi} = 0$, should progressively smooth out, ultimately leading to a vanishing physical mass. A direct numerical integration of the flow equations in this regime would thus provide valuable insight into the mechanism by which scale invariance is restored.

In the same spirit, while our analysis in Sec.~\ref{sec:linearized}
clarified the discontinuous behavior of the correlation-length exponent
$\nu$---which jumps from $1/2$ to $1/3$  $N=\infty$ limit---it
would be highly valuable to investigate at large but finite $N$ the dynamical implications of this regularized singularity by integrating the full nonlinear RG flow. In particular, one may envision a scenario in
which the RG trajectory, although initially governed by a singular
perturbation and therefore characterized by an effective exponent close
to $\nu=1/3$, is progressively influenced by boundary-layer effects and
ultimately crosses over to a scaling regime with $\nu\simeq 1/2$ in the
infrared limit $k\to 0$. Understanding how such a dynamical crossover is
encoded in physical observables would provide valuable insight into the
fate, at large but finite $N$, of the discontinuity that emerges in the
strict $N=\infty$ limit.

Finally, phenomena akin to the Bardeen--Moshe--Bander fixed point in $O(N)$ models have also been reported in fermionic systems, such as Gross--Neveu models \citep{cresswell2023line} and supersymmetric $O(N)$ theories. The construction of singular potentials, as well as their extension to finite $N$ within the Wilson--Polchinski formulation, can be naturally generalized to models involving fermionic degrees of freedom. Pursuing this direction could provide further insight into the mechanisms underlying mass generation in such systems.

\begin{acknowledgments}
This work is supported by JSPS Grant-in-Aid for Scientific Research
(Grants Nos. 21K03488 and 24K06984). 
\end{acknowledgments}

\bibliographystyle{apsrev}
\addcontentsline{toc}{section}{\refname}\bibliography{biblio-3}

\appendix

\end{document}